\documentclass[onecolumn,draftclsnofoot,12pt]{IEEEtran}
\usepackage{amsfonts}
\usepackage{times}
\usepackage{graphicx}
\usepackage{latexsym}
\usepackage{dsfont}
\usepackage{amssymb}
\usepackage{amsmath}
\usepackage{cite}
\usepackage{verbatim}
\usepackage{subfigure}

\newcommand{\figref}[1]{{Fig.}~\ref{#1}}

\def\bb0{{\mathbb{0}}}

\def\ba{{\mathbf{a}}}
\def\bb{{\mathbf{b}}}

\def\bg{{\mathbf{g}}}

\def\b0{{\mathbf{0}}}

\def\bX{{\mathbf{X}}}

\def\sf0{{\mathsf{0}}}

\usepackage{epstopdf}
\usepackage{enumerate}
\usepackage{algorithmicx}
\usepackage{algorithm}
\usepackage{amsmath}
\usepackage{mathtools}
\usepackage[noend]{algpseudocode}
\usepackage{float}
\usepackage{hyperref}
\usepackage{color}
\usepackage{makeidx}
\usepackage{bbm}
\usepackage{graphicx}
\usepackage{lipsum}
\usepackage{subfigure}
\usepackage{tablefootnote}
\usepackage{multirow}
\usepackage{multicol}
\usepackage{balance}
\usepackage{booktabs}
\usepackage{soul}
\usepackage{xcolor}
\usepackage{autobreak}
\usepackage[utf8]{inputenc}
\usepackage{tabularx}
\usepackage{gensymb}

\def\j{\mathrm{j}}

\newcommand{\comm}[1]{}

\def\BibTeX{{\rm B\kern-.05em{\sc i\kern-.025em b}\kern-.08em
		T\kern-.1667em\lower.7ex\hbox{E}\kern-.125emX}}
\AtBeginDocument{\definecolor{ojcolor}{cmyk}{0.93,0.59,0.15,0.02}}

\begin{document}
	
	\title{ Sensing-Aided 6G Drone Communications: Real-World Datasets and Demonstration}
	
	\author{Gouranga Charan and Ahmed Alkhateeb \footnote{The authors are with the School of Electrical, Computer, and Energy Engineering, Arizona State University. Emails: \{gcharan, alkhateeb\}@asu.edu.  This work was supported by the National Science Foundation (NSF) under Grant No. 2048021.  Part of this work has been accepted in
			the IEEE Global Communications Conference \cite{drones_conf}.}}
	
	\maketitle
	
	\begin{abstract}
		In the advent of next-generation wireless communication, millimeter-wave (mmWave) and terahertz (THz) technologies are pivotal for their high data rate capabilities. However, their reliance on large antenna arrays and narrow directive beams for ensuring adequate receive signal power introduces significant beam training overheads. This becomes particularly challenging in supporting highly-mobile applications such as drone communication, where the dynamic nature of drones demands frequent beam alignment to maintain connectivity. Addressing this critical bottleneck, our paper introduces a novel machine learning-based framework that leverages multi-modal sensory data, including visual and positional information, to expedite and refine mmWave/THz beam prediction. Unlike conventional approaches that solely depend on exhaustive beam training methods, our solution incorporates additional layers of contextual data to accurately predict beam directions, significantly mitigating the training overhead. Additionally, our framework is capable of predicting future beam alignments ahead of time. This feature enhances the system's responsiveness and reliability by addressing the challenges posed by the drones' mobility and the computational delays encountered in real-time processing. This capability for advanced beam tracking asserts a critical advancement in maintaining seamless connectivity for highly-mobile drones. We validate our approach through comprehensive evaluations on a unique, real-world mmWave drone communication dataset, which integrates concurrent camera visuals, practical GPS coordinates, and mmWave beam training data. Our findings demonstrate a top-$1$ beam prediction accuracy of $86.32\%$, with near-perfect top-3 and top-5 accuracies, showcasing a substantial reduction in beam training overhead.  For future beam predictions, our vision-aided solution achieves notable accuracies of approximately $92\%$ and $88\%$ for predicting two and three future beams respectively, using the top-3 accuracy metric.  These results not only underscore the efficacy of our sensing-aided solution but also mark a significant stride towards realizing efficient and highly-mobile $6$G drone communications, potentially transforming future aerial networks.
		
	\end{abstract}

	\maketitle

	\section{Introduction} \label{sec:Intro}

	The emergence of highly mobile aerial platforms, particularly drones and unmanned aerial vehicles (UAVs), presents both opportunities and challenges for future wireless communication systems in 5G-advanced, 6G, and beyond \cite{bariah2020drone, mozaffari2021toward}. Drones can act as mobile aerial base stations to extend the coverage of millimeter-wave (mmWave) and sub-terahertz (THz) networks in urban canyons, rural areas, and temporary large-scale events. In disaster scenarios, drones can rapidly establish emergency communication infrastructure in affected areas where traditional networks are compromised \cite{mukherjee2022disastdrone}. For security-critical applications, drones can implement directional beam patterns that minimize interference to legitimate users while creating controlled interference zones to protect against potential eavesdroppers \cite{zhang2024decentralized}, all while maintaining ultra-low-latency requirements. To meet the demanding data rate requirements of these applications, drones must be equipped with mmWave/THz transceivers \cite{Rappaport2019} and large-scale antenna arrays. However, the effective operation of these systems hinges on precise beam alignment between transmitters and receivers to maintain adequate signal-to-noise ratio (SNR). This beam management becomes particularly challenging in drone communications due to two critical factors: first, the training overhead scales significantly with the number of antenna elements, and second, the unique three-dimensional mobility patterns of drones necessitate more frequent beam adjustments compared to traditional ground-based systems. These challenges underscore the need for innovative approaches to enable efficient and reliable mmWave/THz drone communications while minimizing the beam training overhead.
	
	\subsection{Prior Work}

	Several solutions have been developed to address the beamforming and channel estimation overhead \cite{Hur2011,Alkhateeb2014, Jayaprakasam2017,Alkhateeb2018, Khan2020a, Alrabeiah_camera}. Initial approaches followed three main directions: adaptive beam codebook training \cite{Hur2011, Alkhateeb2014}, compressive channel estimation using channel sparsity \cite{Alkhateeb2014}, and beam tracking solutions \cite{Jayaprakasam2017}. The first approach employs adaptive beam codebooks for beam training \cite{Hur2011, Alkhateeb2014}. However, this method requires extensive training time, particularly in mobile multi-user scenarios, where beam patterns must be frequently updated. The second approach exploits the inherent sparsity of mmWave channels to formulate channel estimation as a compressive sensing problem \cite{Alkhateeb2014}. This method reduces the number of required measurements by reconstructing the channel from a small set of observations. However, even with this reduction, the training overhead remains substantial, achieving only one order of magnitude improvement. Additionally, the training overhead scales with the antenna array size, making it hard-to-scale for mmWave/THz wireless communication systems with large antenna arrays. Next, to address the limitations of static approaches, researchers investigated dynamic tracking methods. Specifically, \cite{Jayaprakasam2017} developed an extended Kalman filter-based (EKF) channel tracking framework to maintain the communication link between the basestation and mobile user. While this EKF-based approach reduces beam training overhead, its practical implementation faces two key limitations: the prediction capability is restricted to short time windows, and the performance degrades significantly in NLOS scenarios. These collective limitations of conventional approaches necessitate the development of more efficient beam prediction methods.
	
	The emergence of machine learning (ML) has introduced novel approaches to address the beam prediction challenge \cite{Alkhateeb2018, RobertPos, Elizabeth, charan2021c, Charan2021b, Morais22, Jiang_LiDAR, demirhan2021beam, Charan_ML22}. These methods mainly focus on utilizing additional environmental information to enhance prediction accuracy. Initial ML approaches focused on wireless signatures \cite{Alkhateeb2018} to predict optimal beam indices at the basestation. However, these methods were constrained to single-user scenarios, limiting their practical application. Subsequent research explored position-based prediction methods \cite{RobertPos, Elizabeth, Morais22}. While these approaches demonstrate potential in reducing training overhead, their reliance on position information introduces prediction inaccuracies due to inherent GPS errors. Recent advances in vehicle-to-infrastructure (V2I) and vehicle-to-vehicle (V2V) communications have incorporated additional sensing modalities, including LiDAR, radar, and cameras \cite{charan2021c, Charan2021b, Jiang_LiDAR, demirhan2021beam, Charan_ML22, morais2024deepsense}. These solutions are specifically designed for vehicles operating in a two-dimensional plane with predictable mobility patterns. Drone communications, however, present a fundamentally different challenge. The six degrees of freedom in drone movement—three translations and three rotations—significantly increase the complexity of beam prediction, requiring new approaches beyond those developed for vehicular communications.

	The potential of mmWave/THz drone communication in enabling novel applications has led to increased research interest in this domain. Recent studies have specifically addressed beam management challenges in drone communications \cite{Song2021a, Yuan2020a, Liu2021a}. The distinct mobility characteristics of drones, including their high speed and variable orientation capabilities, introduce unique challenges for beam prediction. These studies examine critical factors affecting beam alignment, such as orientation changes and platform jitter, while developing drone-specific beam prediction solutions. Furthermore, they utilize certain user-side (drone) information to develop solutions to overcome these challenges and accurately predict optimal beams. In \cite{Liu2021a}, the authors propose to use the computed past location to predict the future location of the drone and thereby predict the optimal beam index. The location is calculated based on the previous location and the current speed of the drone, with some added margin for error. This solution, although promising, is based on simulation data. Given the higher speeds of the drone, the error between the predicted and the actual location can vary significantly in a real-world setting. In \cite{Song2021a, Yuan2020a}, the authors proposed utilizing the angle between the basestation and the drone as an input to predict the future beam index. Although these solutions can help reduce the beam training overhead, it might be difficult to acquire the accurate angle between the basestation and drone in a time-efficient manner. Furthermore, all these solutions are designed to predict only for the next time unit by utilizing information available up to the current time step. With the drones' higher speed and the latency associated with data transfer, a more realistic approach will be to predict further into the future. To efficiently achieve this capability, holistic information about the wireless environment is critical.

	\subsection{Contribution}
	Previous works \cite{charan2021c, Charan2021b, morais2024deepsense} have demonstrated promising results in sensing-aided beam prediction for V2I and V2V scenarios. This paper investigates whether similar approaches can be effectively extended to mmWave/THz drone communications. We propose a deep learning framework that leverages multi-modal sensing data to reduce beam training overhead in drone communications. The primary contributions of this work are:
	\begin{itemize}
		\item {A comprehensive formulation of the sensing-aided beam prediction problem for mmWave/THz drone communications. This formulation incorporates practical constraints from both sensing modalities (visual data and other drone features such as orientation, altitude, and GPS coordinates) and communication models, addressing the unique challenges of aerial platforms.}
		
		\item{Development of a novel deep learning architecture that integrates multiple sensing modalities for both current and future beam prediction. The proposed solution combines visual data captured at the basestation with drone telemetry data, including position coordinates, orientation parameters, and altitude information to predict the optimal beams.}
		
		\item{Development of the drone communication scenario (Scenario 23) in DeepSense 6G \cite{DeepSense}, a large-scale dataset that synchronously captures visual data and drone navigation parameters along with the corresponding mmWave beam indices.}
			
		\item{Comprehensive experimental validation using the DeepSense 6G dataset, representing the first real-world evaluation of sensing-aided beam prediction and tracking for drone communications. The results demonstrate the practical viability of multi-modal sensing for beam prediction in aerial platforms.}
		
	\end{itemize}

	\section{Sensing-Aided 6G Drone Communications: \\ Beam Prediction and Tracking}\label{sec:pred_key_idea}

	Addressing the challenge of beam training overhead is of paramount importance in enabling high-frequency mmWave/THz drone wireless communication systems. Recent work has shown interest in leveraging artificial intelligence and deep neural networks to tackle the beam training overhead challenge. One of the critical directions involves utilizing additional sensory data such as vision, LiDAR, Radar, and GPS positions, to name a few, in order to reduce the beam training overhead. This work uniquely focuses on leveraging vision and positional data to perform two important tasks: (i) current beam prediction and (ii) future beam tracking. In the next two subsections, we present the motivation and the key idea behind our sensing-aided beam prediction and tracking task.

	\begin{figure}[!t]
		\centering
		\includegraphics[width=0.8\linewidth]{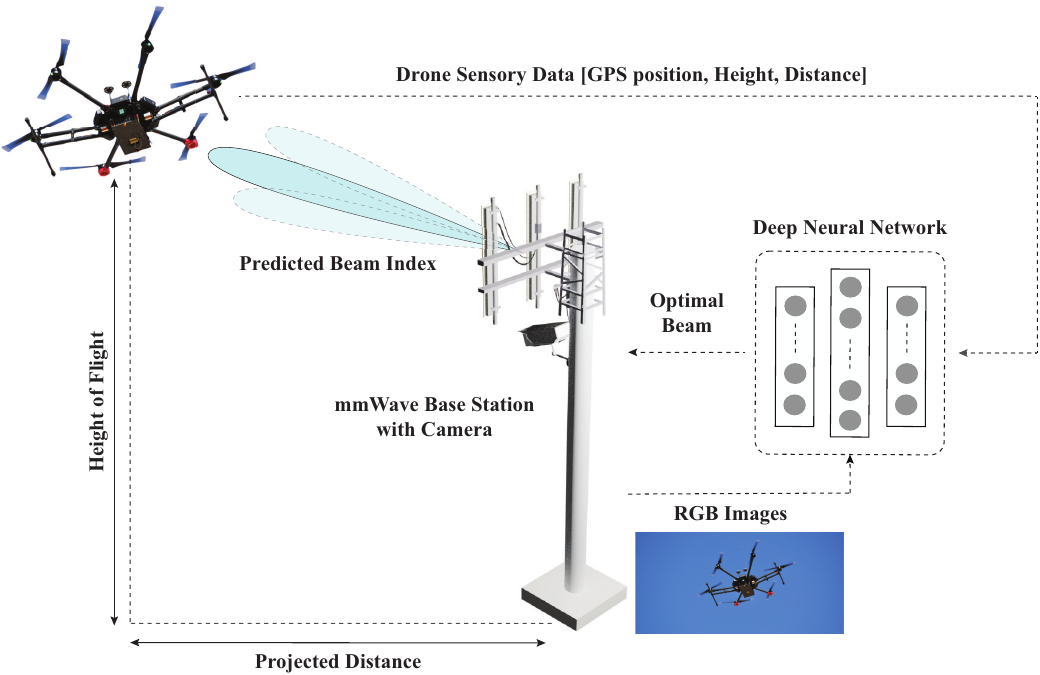}
		\caption{An illustration of the mmWave base station serving a drone in a real wireless environment. The base station utilizes additional sensing data, such as RGB images and the GPS location of the drone, to predict the optimal beam indices.}
		\label{fig:beam_pred_key_idea}
	\end{figure}
	
	\subsection{Sensing-Aided mmWave Drone Beam Prediction}
	
	The mmWave/THz communication systems employ large antenna arrays with narrow directive beams to compensate for the severe path loss inherent at high frequencies. The selection of optimal beams, however, incurs significant training overhead. This challenge is particularly pronounced in dynamic environments with mobile transmitters and varying channel conditions. The complexity further increases in drone communications due to their unique mobility characteristics: high flying speeds, ability to hover, and freedom to traverse three-dimensional space. To address these challenges, we propose a sensing-aided approach that leverages additional sensory data for optimal beam selection. Specifically, we formulate beam selection as a prediction task where the optimal beam index is chosen from a pre-defined codebook at each coherence time.
	
	The fundamental premise of our sensing-aided solution stems from the predominant line-of-sight (LOS) nature of high-frequency communications. In well-calibrated antenna arrays, beamforming vectors provide directional information that characterizes the dominant signal paths. These vectors partition the spatial dimensions into multiple, potentially overlapping sectors, with each sector corresponding to a specific beam index. This spatial partitioning allows us to reformulate the beam prediction task as a classification problem, where beam indices are assigned based on the user's position in the wireless environment. Our approach leverages two key technological advances: state-of-the-art computer vision systems that enable precise object detection and spatial relationship extraction, and advanced positioning systems that provide accurate location information with quantifiable error margins. As illustrated in Fig.~\ref{fig:beam_pred_key_idea}, our system comprises a high-frequency base station serving a drone equipped with a mmWave transmitter. The base station integrates a well-calibrated high-frequency antenna array with an RGB camera to capture environmental visual data at each time step. A machine learning framework processes this multi-modal sensory data to predict the optimal beamforming vector from the pre-defined codebook, thereby eliminating the need for continuous beam training. The detailed architecture and implementation of the proposed solution are presented in Section~\ref{sec:beam_pred}.

	\subsection{Sensing-Aided mmWave Drone Beam Tracking}
	
	In the previous subsection, we presented the premise and the motivation behind sensing-aided mmWave drone beam prediction. Here, we focus on our next objective, which is sensing-aided mmWave drone beam tracking. The critical question we address is, \textit{what distinguishes beam tracking from the beam prediction task?} The sensing-aided beam prediction task involves predicting the current optimal beam index based on the available sensory data at any given time instant. However, there is a trade-off between the reduced beam training overhead and the computational cost. The adoption of deep neural networks, while beneficial for enhancing prediction accuracy, significantly contributes to this computational overhead due to their complex processing requirements. The computational complexities of the beam prediction solutions result in increased latency costs associated with these approaches. The sensory data acquisition and transfer between nodes are also time-intensive processes and further add to the latency cost of the sensing-aided beam prediction solutions. As such, the added latency cost makes it difficult for these solutions to satisfy the ultra-low reliability requirements of several future applications. An efficient solution to overcome latency-related challenges can be proactively predicting future beams. The proactive prediction of future beams helps in a partial or complete offset of the increased prediction latency.
	
\begin{figure}[!t]
		\centering
		\includegraphics[width=0.9\linewidth]{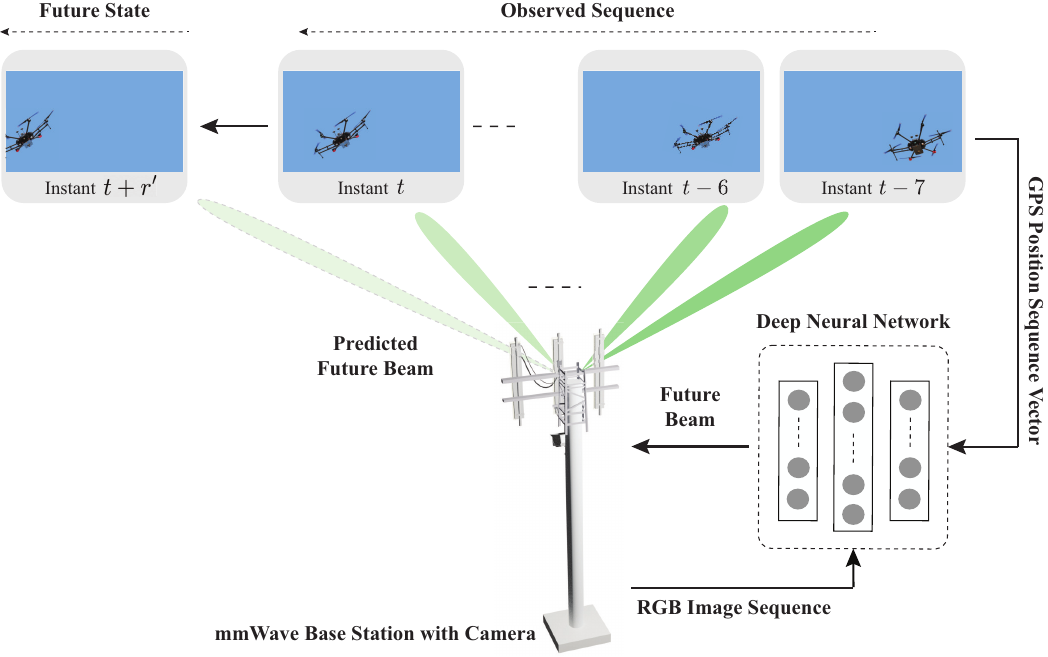}
		\caption{An illustration to highlight the future beam prediction/beam tracking task. The mmWave base station utilizes a sequence of past sensing data such as RGB images ans GPS locations of the transmitter to predict the future optimal beam indices. }
		\label{fig:beam_track_key_idea}
	\end{figure}

	The beam tracking task involves predicting the optimal future beams by utilizing current and previously observed information (for example, a sequence of images). In this work, we focus on predicting up to three future beams using the available sensing information. The dynamic nature of the wireless environment makes the future beam-tracking task far more challenging than predicting the current beam. Gaining a comprehensive understanding of the wireless environment becomes pivotal in predicting future beams with high accuracy. The available sensing information, such as the visual and position data, can help provide relevant information about the transmitter and the wireless environment in general, further highlighting the potential of leveraging additional sensing data for such tasks. Apart from aiding in the wireless scene understanding, the sequence of visual or position data can also help in providing crucial information about the transmitter, such as the speed of motion, the direction of travel, the location of the user in the wireless scene, etc. All this information, combined with the knowledge of the environment, plays a central role in facilitating future beam prediction or beam tracking tasks. Fig.~\ref{fig:beam_track_key_idea} highlights the key idea of the future beam-tracking task. In order to predict the future optimal beams, the base station observes a sequence of current and previously observed sensing data. The preferred sensing data for this task are the sequence of past GPS locations of the drone and the sequence of RGB images. The base station is then proposed to predict the future beams using this sequence data as the input. The detailed proposed solution will be presented in Section~\ref{sec:beam_track}.
	
	\section{System Model}\label{sec:sys_mod}
	
	The proposed sensing-aided drone communication framework adopts a system architecture illustrated in Fig.~\ref{fig:beam_pred_key_idea} and Fig.~\ref{fig:beam_track_key_idea}. The system consists of a basestation equipped with two key components: a uniform linear array (ULA) of $M$ elements for communication and an RGB camera for visual sensing. The basestation serves a drone that functions as a single-antenna transmitter and incorporates a GPS receiver for real-time position tracking. For beamforming operations, the basestation utilizes a pre-defined codebook denoted as $\boldsymbol{\mathcal F}=\{\mathbf f_q\}_{q=1}^{Q}$, where $\mathbf{f}_q \in \mathbb C^{M\times 1}$ represents individual beamforming vectors, and $Q$ denotes the total number of available beamforming vectors. The mmWave communication system implements OFDM with $K$ subcarriers and employs a cyclic prefix of length $D$. At time step $t$, the channel between the basestation and the drone at the $k$th subcarrier is represented by $\mathbf h_{k}[t] \in \mathbb C^{M\times 1}$. For uplink transmission, when the drone transmits a complex symbol $x\in \mathbb C$, the received signal $y_{k}[t] \in \mathbb C$ at the base station for the $k$th subcarrier is expressed as:
	
	\begin{equation}\label{eq:sys_mod}
		y_{k}[t] = \mathbf h_{k}^T[t] \mathbf f_q[t]x + \beta_k[t],
	\end{equation}
	
	where $\mathbf f \in \boldsymbol{\mathcal F}$ represents the optimal beamforming vector at time $t$, and $\beta_k[t]$ denotes the complex Gaussian noise following distribution $\mathcal N_\mathbb C(0,\sigma^2)$. The transmitted symbol $x\in \mathbb C$ must satisfy the power constraint $\mathbb E\left[ |x|^2 \right] = P$, where $P$ defines the power budget per symbol. The optimal beamforming vector $\mathbf f^{\star}[t]$ at each time step $t$ is determined through classical beam training, which involves sequential evaluation of vectors from codebook $\mathcal F$. This beam sweeping process can be mathematically formulated as
	\begin{equation}\label{eq:beam_training}
		\mathbf f^{\star}[t] = \underset{\mathbf f_q[t]\in \mathcal F}{\text{argmax}} \frac{1}{K}\sum_{k=1}^{K} \mathsf{SNR}|\mathbf h_{k}^T[t] \mathbf f_q[t] |^2,
	\end{equation}
	where $\mathrm{SNR}$ is the signal-to-noise ratio. A geometric mmWave channel model is adopted in this work. The channel vector $\mathbf h_k$ in \eqref{eq:sys_mod} at the $kth$ subcarrier  is given by
	
	\begin{equation}\label{eq:channel_mod}
		\mathbf{h}_{k} = \sum_{d=0}^{D-1} \sum_{\ell=1}^L \alpha_\ell e^{- \j \frac{2 \pi k}{K} d} p\left(dT_\mathrm{S} - \tau_\ell\right) \ba\left(\theta_\ell, \phi_\ell\right),
	\end{equation} 
	where $L$ is a number of channel paths, $\alpha_\ell, \tau_\ell, \theta_\ell, \phi_\ell$ is the path gains (including the path-loss), the delay, the azimuth angle of arrival, and the elevation angle of arrival, respectively, of the $\ell$th channel path. $T_\mathrm{S}$ represents the sampling time while $D$ denotes the cyclic prefix length (assuming that the maximum delay is less than $D T_\mathrm{S}$). Considering a block-fading channel model, the channel is assumed to stay constant over the channel coherence time.

	%%%%%%%%%%%%%%%%%%%%%%%%%%%%%%%%%%%%%%%%%%%%%%%%
	
	\section{Sensing-Aided Beam Prediction for mmWave Drones}\label{sec:beam_pred}
	This section studies the sensing-aided beam prediction problem in a real-wireless setting. This section is divided into three parts. The first subsection includes the formal definition of the sensing-aided beam prediction task. The second subsection formally defines the machine learning task for both position and vision-aided beam prediction tasks. The final subsection includes the proposed deep learning-based solution to the current problem.

	\subsection{Problem Formulation}\label{sec:beam_pred_prob_form}

	Based on the system model presented in Section~\ref{sec:sys_mod}, the optimal beam selection problem at the base station involves choosing a beam $\mathbf f^{\star}[t]$ from its codebook $\boldsymbol{\mathcal F}$ to maximize the received power at the drone, as defined in \eqref{eq:beam_training}. Conventional approaches rely on explicit channel knowledge $\mathbf h_{k}$ to determine the optimal beam. However, acquiring accurate channel state information in mmWave/THz systems is particularly challenging due to the large antenna arrays and the highly dynamic nature of drone communications. To address this challenge, we propose an alternative approach that leverages multi-modal sensory data to predict the optimal beam index, thereby circumventing the need for explicit channel estimation. The proposed framework utilizes three categories of sensory data collected at the base station and the drone. RGB images $\bX[t] \in \mathbb{R}^{W \times H \times C}$ are captured using a camera installed at the base station at time $t$, where $W$, $H$, and $C$ represent the image width, height, and number of color channels, respectively. The drone's GPS positional data, represented as a two-dimensional vector $\bg[t] \in \mathbb R^2$ containing latitude and longitude coordinates at time $t$, is fed back to the base station. Additionally, spatial parameters including the height $d[t] \in \mathbb R^1$ and distance $v[t] \in \mathbb R^1$ of the drone relative to the base station at time $t$ provide crucial three-dimensional positioning information in the real wireless environment. The beam prediction task can thus be formulated as determining a mapping function that processes the available sensory data subset $\mathcal{S}[t] \subset \left\lbrace \bg[t], \bX[t], d[t], v[t]\right\rbrace$ to predict the optimal beam $\mathbf f^{\star}[t]$ with high accuracy. This formulation transforms the traditional channel estimation problem into a learning-based prediction task that leverages available sensory information.

	\subsection{Machine Learning Task}\label{sec:beam_pred_ml_def}
	Given the objective in Section~\ref{sec:beam_pred}-\ref{sec:beam_pred_prob_form}, in this work, we target developing machine learning (ML) models to learn the beam prediction mapping function. This work focuses on presenting a comprehensive evaluation of both position and vision-aided mmWave drone beam prediction. In order to do so, both sub-tasks need to be first defined formally. We start this subsection by defining the ML task for position-aided beam prediction, followed by a vision-aided beam prediction problem statement. 
	
	\subsubsection{Position-Aided Drone Beam Prediction}\label{sec:pos_beam_pred_ML}
	Formally, the ML task for position-aided beam prediction in drone mmWave communication can be defined as follows. 
	\begin{quote}
		\textit{Position-aided drone beam prediction is the task of predicting the optimal beam indices from a predefined codebook with the real-time position of the drone in the real-wireless environment. Given a dataset of GPS coordinates and corresponding optimal beam indices, the goal is to train a model that accurately predicts the beam index based on the real-time position of the drone in a wireless environment. }
	\end{quote} 
	
	The machine learning task utilizes a dataset $ \mathcal D_{\text{task}_{1-1}} = \left\lbrace \left (\bg_u, \mathbf f^{\ast}_u \right) \right\rbrace_{u=1}^U $ acquired from a real wireless environment, where $U$ denotes the total number of samples. Each data point comprises a pair of GPS position coordinates and its corresponding optimal beam vector. The objective is to develop a machine learning model that learns a prediction function $f_{\Theta_1}(\bg_u)$, which maps observed positions to probability distributions $\mathcal P \in \{p_1, \ldots, p_{Q} \}$ over the beamforming codebook vectors $\boldsymbol{\mathcal F}$. The prediction function is parameterized by a set $\Theta_1$, representing the model parameters that are optimized using the labeled dataset $ \mathcal D_{\text{task}_{1-1}}$. The primary objective is to maximize prediction accuracy across all samples in the dataset. Under the assumption that the samples in $ \mathcal D_{\text{task}_{1-1}}$ are independent and identically distributed (i.i.d.), this optimization problem can be formally expressed as:
	\begin{equation}\label{eq:prob_form_1}
		f^{\star}_{\Theta_1^{\star}} = \underset{f_{\Theta_1}}{\text{argmax}}\\ \prod_{u=1}^{U_1} \mathbb P\left( \hat{\mathbf f}_u = \mathbf f^{\star}_u | \mathcal{S}_u \right),
	\end{equation}
	where the product formulation in \eqref{eq:prob_form_1} follows directly from the i.i.d. assumption of the dataset samples.

	Given the six degrees of freedom in drones, other sensory information, such as the drone's height and distance, might play an important role in the beam prediction task. Therefore, the in-depth study of the drone beam prediction task necessitates considering the other sensing modalities. In order to study the impact of the other modalities, we first develop a dataset $ \mathcal D_{\text{task}_{1-2}} = \left\lbrace \left (\mathcal S^{\star}_u , \mathbf f^{\ast}_u \right) \right\rbrace_{u=1}^{U{_2}} $, where $\mathcal S^{\star}_u  \subset \left\lbrace \bg_u, d_u, v_u\right\rbrace$ is a tuple of multimodal data samples that include GPS position, the height and distance of the drone, and $U_2$ is the total number of samples in this dataset $ \mathcal D_{\text{task}_{1-2}}$. Similar to the position-aided beam prediction solution, the ML model is developed to learn a prediction function $f_{\Theta_2}(\mathcal S^{\star}_u)$, where $\Theta_2$ represent the model parameters. The function needs to maximize the probability of correct beam predictions given the position, height, and distance data, i.e.,
	\begin{equation}\label{eq:prob_form_2}
		f^{\star}_{\Theta_2^{\star}} = \underset{f_{\Theta_2}}{\text{argmax}}\\ \prod_{u=1}^{U_2} \mathbb P\left( \hat{\mathbf f}_u = \mathbf f^{\star}_u | \mathcal S^{\star}_u \right),
	\end{equation} 
	Similar to the \eqref{eq:prob_form_1}, the product in \eqref{eq:prob_form_2} is a result of an implicit assumption that the samples of $ \mathcal D_{\text{task}_{1-2}}$ are of i.i.d. nature.

	\subsubsection{Vision-Aided Drone Beam Prediction}\label{sec:img_beam_pred_ML}
	In the previous task, we propose to utilize positional and other relevant sensing information, such as the height and distance of the drone, to predict the optimal beams. Although these modalities can capture important information regarding the drone's location in the wireless environment, which is essential for the beam prediction task, they are often faced with several challenges. One of the significant challenges is the error in GPS positional data. Several factors, such as atmospheric delay, receiver noise, etc., impact the precision of the received position data. Therefore, it is interesting to investigate if other sensing modalities can be utilized to reduce the beamforming overhead. One exciting direction that has recently shown promise is the visual data captured by the base station. The second approach we study in this paper is the ML-based vision-aided mmWave drone beam prediction. The ML problem statement for this task can be defined as follows.
	
	\begin{quote}
		\textit{This task aims to predict the optimal beam index using images captured by a base station camera. The model learns from a dataset comprising RGB images and their associated optimal beam indices, with the objective to accurately determine the beam index from visual data. }
		
	\end{quote} 
	
	The task is formally defined similarly to the position-aided beam prediction problem. The vision-aided beam prediction task involves predicting the optimal beamforming vector based on the image captured by the base station at any given time. The beam prediction task is formulated as an optimization problem, where the function is learned using an ML model. This prediction function is parameterized by a set $\Theta_3$ representing the ML model parameters and learned from a dataset of labeled samples, i.e., $ \mathcal D_{\text{task}_{1-3}} = \left\lbrace (\b{X}_{u}, \mathbf f^{\ast}_u) \right\rbrace_{u=1}^{U_3} $, where $U_3$ is the total number of samples in the dataset, $\b{X}_{u}$ is the RGB input image captured by the camera installed at the base station and $\mathbf f^{\ast}_u$ is the optimal beam vector in ${\mathcal F}$ and associated with the $u$th sample in the dataset. The goal of the function, therefore, is to maximize the probability of correct prediction over the entire dataset. Formally, this can be expressed as
	
	\begin{equation}\label{eq:prob_form_3}
		f^{\star}_{\Theta_3^{\star}} = \underset{f_{\Theta_3}}{\text{argmax}}\\ \prod_{u=1}^{U_3} \mathbb P\left( \hat{\mathbf f}_u = \mathbf f^{\star}_u | \bX_u \right),
	\end{equation}
	where the samples in dataset $\mathcal D_{\text{task}_{1-3}}$ are independent and identically distributed (i.i.d.). 
	
	\begin{figure}[!t]
		\centering
		\includegraphics[width=1.0\linewidth]{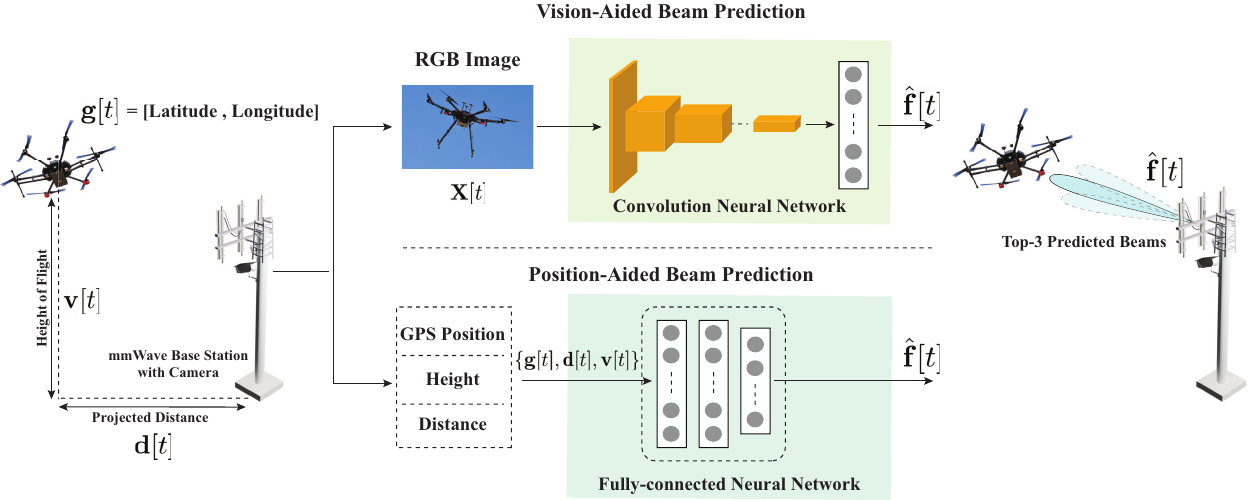}
		\caption{A block diagram showing the proposed solution for both the vision and position-aided beam prediction task. As shown in the figure, the camera installed at the base station captures real-time images of the drone in the wireless environment. A CNN is then utilized to predict the optimal beam index. For the other three sensing data, the base station receives the information, which is then passed through a fully connected neural network to predict the beam. }
		\label{fig:beam_pred_soln}
	\end{figure}

	\subsection{Proposed Sensing-Aided Beam Prediction Solution}\label{sec:beam_pred_prop_soln}
	Building upon the machine learning tasks defined in Section~\ref{sec:beam_pred}-\ref{sec:beam_pred_ml_def}, this section presents a detailed analysis of the proposed sensing-aided beam prediction solutions. The organization is structured as follows: first, the position-aided beam prediction approach is presented, followed by the vision-aided beam prediction methodology. The block diagram of the proposed beam prediction ML model is illustrated in Fig.~\ref{fig:beam_pred_soln}.
	
	\subsubsection{Proposed Position-aided Solution} \label{sec:position_beam_pred}
	The position-aided approach focuses on predicting the optimal beam index utilizing the transmitter's positional information. The architecture employs a Multi-Layer Perceptron (MLP) network that processes normalized Latitude and Longitude values as input features. The designed MLP architecture consists of two hidden layers, each comprising $512$ hidden units, followed by an output layer with $Q$ units. Non-linearity is introduced through ReLU activation functions applied to the hidden layer outputs. Given that the position-aided beam prediction is formulated as a classification problem in Section~\ref{sec:beam_pred}\ref{sec:beam_pred_ml_def}-\ref{sec:pos_beam_pred_ML}, a softmax function is implemented at the final output layer. To enable comprehensive evaluation across sensing modalities, this architecture is enhanced to incorporate additional spatial features - specifically, the normalized height and distance measurements between the drone and the base station. This augmented version maintains the core architectural elements while expanding the input feature set beyond GPS coordinates.
	
	\subsubsection{Proposed Vision-aided Solution} \label{sec:vision_beam_pred}
	This section details the deep learning architecture developed for vision-aided beam prediction. Following the formulation in Section~\ref{sec:beam_pred}\ref{sec:beam_pred_ml_def}-\ref{sec:img_beam_pred_ML}, the objective involves learning the prediction function $f_{\Theta}(\bX[t])$ utilizing RGB images as the sole input. Convolutional Neural Networks (CNNs) emerge as the natural architectural choice for this image classification task. The selected CNN architecture addresses two critical requirements: achieving high classification accuracy with robust generalization capabilities, while maintaining minimal memory footprint and computational latency. The residual neural network (ResNet) \cite{resnet} architecture, specifically ResNet-50, is selected for its demonstrated efficiency in meeting these requirements. Rather than training from scratch, the implementation begins with a ResNet model pre-trained on ImageNet2012 \cite{ImageNet12}. This base architecture undergoes modification through the replacement of its final fully connected layer with a new layer containing $Q$ neurons, initialized using a normal distribution (zero mean, unit variance). The training methodology diverges from traditional transfer learning approaches \cite{transfer_learning}, implementing end-to-end fine-tuning in a supervised manner using the labeled dataset.
	
	%%%%%%%%%%%%%%%%%%%%%%%%%%%%%%%%%%%%%%%%%%%%%%%%
	
	%%%%%%%%%%%%%%%%%%%%%%%%%%%%%%%%%%%%%%%%%%%%%%%%
	\section{Sensing-Aided Beam Tracking for mmWave Drones}\label{sec:beam_track}
	The first phase of the study, as presented in Section~\ref{sec:beam_pred}, dealt with predicting the optimal beam based on the available sensing information at the same time step.  The sensing data explored for the beam prediction task included GPS position, height, and distance of the drone and the RGB images captured by the camera installed at the base station.  The second phase of the study deals with a more challenging problem statement of predicting the future beams at any time instant $t$ using the sensing data available only up to the current time step.  Instead of using the current data sample collected at time $t$ for the beam prediction task, a sequence of samples observed over a time interval of $r$ instances is utilized for beam tracking.  Apart from the visual and position data, we also study the beam-only approach, where the future beams are predicted based on the sequence of past beams.  The performance of the beam-only approach forms the baseline of the beam-tracking task.  Similar to Section~\ref{sec:beam_pred}, this section is divided into three subsections.  The first two subsections present the formal definition of the beam tracking task and define the machine learning task for all three modalities, respectively.  The third subsection presents the proposed solution for all three sensing modalities.
	
	\subsection{Problem Formulation}\label{sec:beam_track_prob_form}
	
	The primary objective of the beam tracking task is still selecting the optimal beam vector $\mathbf f^{\ast}$ from the pre-defined codebook  $ \boldsymbol{\mathcal F}$.  However, instead of predicting the optimal beam for the current time step, we predict the future beam indices.  The objective of the beam tracking task is to utilize sequences of sensing data and develop an optimization algorithm that learns to predict future beam indices.  Formally, this learning could be posed as follows.  At any time instance $\tau\in \mathbb Z$, that sequence is given by
	\begin{equation}\label{eq:beam_track_prob}
		{\mathcal V[\tau]} = \{ (\mathcal{S}[t]) \}_{t = \tau-r+1}^{\tau},
	\end{equation}
	where $\mathcal S[t]$ is a tuple consisting of the GPS position of the drone ($\bg[t] \in \mathbb R^2 $) and the RGB image ($\bX[t] \in \mathbb{R}^{W \times H \times C}$) captured by the base station at $t$th time instant and $r \in \mathbb Z$ is the window of observation. The objective is to observe $\mathcal V$ and predict the future optimal beam indices $\{ \hat{\mathbf{f}}[t]\}_{t=\tau+1}^{\tau+r^\prime}$ in $\mathcal F$ within a window of $r^{\prime}\in \mathbb Z$, where $ t \in \left\lbrace \tau + 1, \ldots, \tau + r^{\prime} \right\rbrace $.

	\subsection{Machine Learning Task}\label{sec:beam_track_prop_soln}
	
	The primary objective of predicting the future beams in the beam prediction task is attained using a machine learning model.  It is developed to learn a prediction function that takes in the observed sensing data and predicts the future beams within a window of $r^\prime$.  From an ML perspective, the beam tracking task can be formally defined as follows. 
	\begin{quote}
		\textit{Sensing-aided drone beam tracking is the task of predicting the future beam indices from a pre-defined codebook by utilizing a machine learning model and sequence of sensory data.  Given a dataset consisting of a sequence of past observations and its corresponding future ground-truth optimal beams, the objective of the ML task is to learn a function that can help predict the optimal future beam indices with high fidelity. }
	\end{quote} 
	
	In this subsection, we formally define the ML task for all three approaches: (i) Beam-only or the baseline method, (ii) position-aided approach, and (iii) vision-aided beam tracking.
	
	\subsubsection{Baseline Approach: Beam-based Future Beam Tracking}\label{sec:baseline_beam_track}
	As described in Section~\ref{sec:pred_key_idea}, the optimal beamforming vector provides directional information.  Observing a sequence of beam indices over a specific window can help understand the drone's motion pattern.  Proactively comprehending where the transmitter is headed is a piece of vital information in predicting future beams successfully.  In this paper, we design an ML model that predicts future beams by utilizing the previously used beams.  In order to do so, a labeled dataset $\mathcal D_{\text{task}_{2-1}} = \left\lbrace (\mathbf f^{\ast}_{u}[t_1], \mathbf b_u[t_2]) \right\rbrace_{u=1}^{U_4} $ consisting of the observed $\mathbf f^{\ast}_{u}[t_1]$ and future optimal beam indices $\mathbf b_u[t_2]$ is collected from a real wireless environment, where $t_1 \in \{\tau -r +1,\ldots, \tau\}$, $t_2 \in \{\tau + 1, \ldots, \tau + r^\prime\}$ and $U_4$ is the total number of samples in the sequence dataset. The ML model is developed to learn a prediction function $f_{\Theta_4}(\mathbf f^{\ast}[t_1])$ that takes in the sequence of observed beam indices and produces a prediction of the future beam values $\mathbf b[t_2] \in \mathcal F$.  The prediction function is parameterized by a set $\theta_4$ representing the model parameters and learned from a dataset of labeled sequences $\mathcal D_{\text{task}_{2-1}}$.  The dataset is then used to train the prediction function $f_{\Theta_4}$ such that the probability of correct prediction over the entire dataset is maximized.  This can be formally expressed as
	\begin{equation}\label{eq:prob_form_4}
		f^{\star}_{\Theta_4^{\star}}  = \underset{f_{\Theta_4}(\mathbf f^{\ast}[t_1])}{\text{argmax}}\\ \prod_{u=1}^{U_4} \mathbb P\left( \hat{\mathbf{f}}_u[t_2] = \mathbf b_u[t_2] \  | \ \mathbf f^{\ast}_u[t_1] \right),
	\end{equation}
	where the joint probability is factored out due to the i.i.d.  nature of the dataset.  The results of the beam-only approach form the baseline results for the beam-tracking task.

	\subsubsection{Position-aided Future Beam Tracking}\label{sec:pos_aided_beam_track_ML}

	The baseline approach presented in Section~\ref{sec:beam_track}\ref{sec:beam_track_prop_soln}-\ref{sec:baseline_beam_track} approaches the beam tracking task from the wireless perspective, where the future beams are predicted based on a sequence of current beams.  Knowledge about the sequence of current beams is a prerequisite for the beam-only approach.  This precondition means that the baseline approach needs to perform beam training to acquire the sequence of current beams, typically associated with high overhead in the mmWave drone communication system.  This work explores the possibility of predicting the future beam using the real-time position information of drones.  Like a sequence of beams, the sequence of positional data can provide meaningful information regarding the drone's location in the wireless environment, the speed, and the direction of travel.  With this motivation, we propose an ML model to perform the position-aided beam-tracking task.  Similar to all the other tasks, we first collect a dataset $\mathcal D_{\text{task}_{2-2}} = \left\lbrace (\bg[t_1], \mathbf b_u[t_2]) \right\rbrace_{u=1}^{U_5} $ consisting of the sequence of $r$ GPS positions $\bg[t_1]$ and $r^\prime$ future beam indices $\mathbf b_u[t_2]$.  $U_5$ denotes the total number of samples in the sequence dataset.  The objective is to learn a prediction function $f_{\Theta_5}(\bg_u)$ by training over the samples in the dataset $\mathcal D_{\text{task}_{2-2}}$, where $\Theta_5$ represents the parameters of the trained ML model.  The goal is to maximize the future beam predictions, which can be formally presented as
	
	\begin{equation}\label{eq:prob_form_5}
		f^{\star}_{\Theta_5^{\star}}  = \underset{f_{\Theta_5}(\bg[t_1])}{\text{argmax}}\\ \prod_{u=1}^{U_5} \mathbb P\left( \hat{\mathbf{f}}_u[t_2] = \mathbf b_u[t_2] \ | \ \bg_u[t_1] \right),
	\end{equation}
	
	where $\hat{\mathbf{f}}_u[t_2]$ is a sequence of $r^\prime$ predicted future beams.

	\subsubsection{Vision-aided Future Beam Tracking}
	
	In Section~\ref{sec:beam_track}\ref{sec:beam_track_prop_soln}-\ref{sec:pos_aided_beam_track_ML}, we formally defined the ML task for predicting the future beams using the real-time GPS position of the user.  The other sensing information of choice is the sequence of RGB images of the drone captured by the camera installed at the base station. One of the concerns with GPS position is the accuracy of the received data.  Any error in the positional data will eventually impact the beam tracking accuracy.  With this motivation, this work studies the vision-aided beam tracking task.  From an ML perspective, the prediction task remains similar to the position-aided beam tracking task as discussed in Section~\ref{sec:beam_track}\ref{sec:beam_track_prop_soln}-\ref{sec:pos_aided_beam_track_ML}.  Given a labeled dataset of RGB images $\bX$ and the sequence of $r^\prime$ future beams $\mathbf b$, this task aims to train an ML model to maximize the correct predictions.  Formally, this can be expressed as
	
	\begin{equation}\label{eq:prob_form_6}
		f^{\star}_{\Theta_6^{\star}}  = \underset{f_{\Theta_6}(\bX[t_1])}{\text{argmax}}\\ \prod_{u=1}^{U_6} \mathbb P\left( \hat{\mathbf{f}}_u[t_2] = \mathbf b_u[t_2] \ | \ \bX_u[t_1] \right),
	\end{equation}
	where $\hat{\mathbf{f}}_u[t_2]$ and $\mathbf b_u[t_2]$ are the predicted future beams and the ground-truth beams, respectively.

	\subsection{Proposed Sensing-Aided Beam Tracking Solution}
\begin{figure}[!t]
		\centering
		\includegraphics[width=1.0\linewidth]{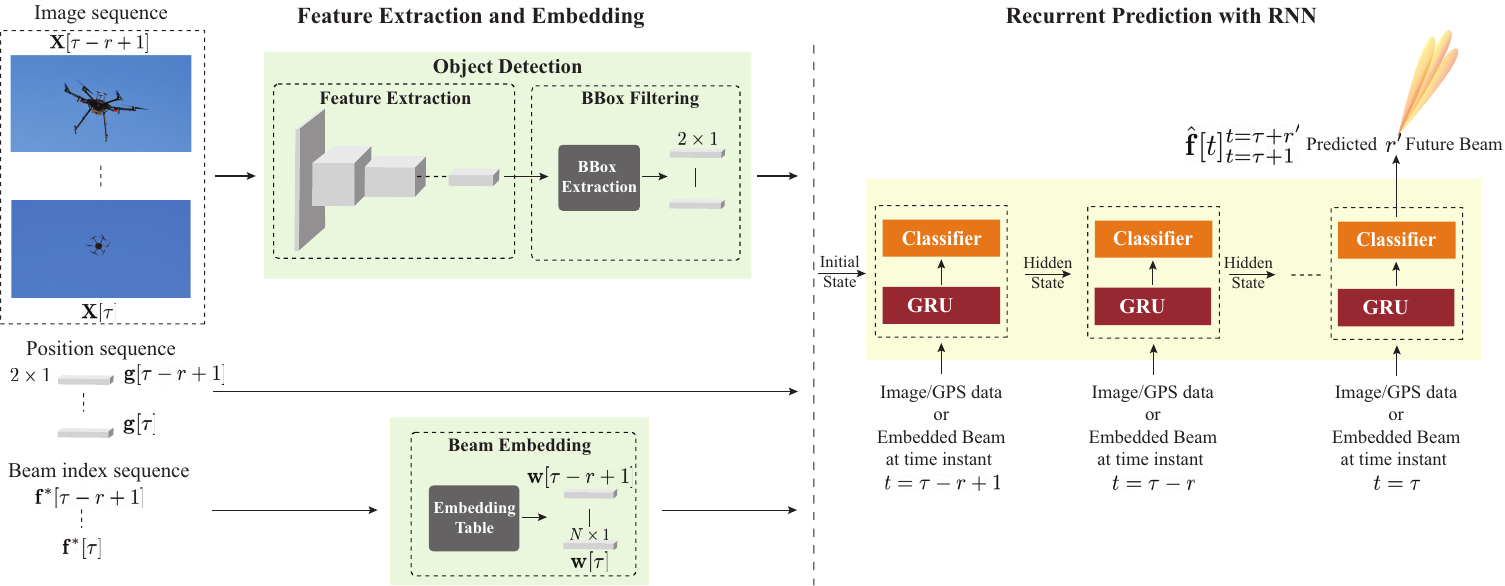}
		\caption{A block diagram showing the proposed solution for the beam tracking task.  It highlights the three different approaches for image, position, and beam sequences.  There are two main stages in the proposed solution: (i) Feature extraction and embedding: For image sequences, this stage is responsible for extracting the feature vectors consisting of the bounding box coordinates of the transmitter and for the beam-based solution this stage embeds the input beam sequences into vectors of dimension $N \times 1$, and (ii) Recurrent prediction with RNN: this stage takes the embedded beam vectors, the feature extracted from the images and the GPS positional data to predict the future beams.   }
		\label{fig:beam_track_soln}
	\end{figure}

	This subsection presents the proposed solution for sensing-aided beam tracking in a real-wireless environment with mmWave drones.  Recurrent neural networks (RNNs) can maintain an internal memory, making them perfectly suited for machine learning problems involving sequential data.  Previously, RNNs have successfully extracted meaningful features in different problems dealing with sequential data, such as natural language processing (NLP) tasks, speech recognition, video activity recognition, etc.  Since the model is expected to learn from a sequence of observations in the beam tracking task, we utilize RNNs and, more specifically, a Gated recurrent unit (GRU) to extract the sequential features.  This study entails three different sensing modalities for the beam tracking task, i.e., GPS position, RGB images, and the wireless mmWave beams.  The inherent differences in the data modalities necessitate three different GRU-based solutions for the beam-tracking task.  The discussion in this subsection is organized as follows.  We first present the details of our deep learning-based solution for the baseline approach.  Next, we present the two different solutions for the position and vision-aided beam tracking task.  Fig.~\ref{fig:beam_track_soln} shows the block diagram with three different solutions for the image, position, and baseline beam tracking task.

	\subsubsection{Baseline Solution} \label{sec:baseline_beam_track_sol}
	The baseline approach aims to predict the future optimal beam indices based on the sequence of observed $r$ previous beams.  RNNs are successful in capturing the sequential dependencies in the input data.  In order to predict the future beams, it is necessary to learn the hidden relation within a sequence of input data.  Therefore, the baseline approach adopts a GRU-based solution to perform the beam-tracking task.  This solution has two major stages: (i) Beam Embedding and (ii) Recurrent Prediction.  Next, we will present a detailed overview of the two stages of the baseline approach.

	\textbf{Beam Embedding:} The first stage of the proposed beam-only solution embeds the raw beam indices into a sparse vector representation.  The adopted beam embedding stage is simple and does not require any training.  It generates a lookup table of $\left|\boldsymbol{\mathcal F}\right|$ real-valued vectors $\mathbf w[t]\in \mathbb R^{N \times 1}$ where $t\in\{\tau-r+1, \dots, \tau\}$. The elements of each vector are randomly drawn from a Gaussian distribution with zeros mean and unity standard deviation.
	
	\textbf{Recurrent Prediction:}
	The second stage of the beam tracking solution consists of an RNN block.  In particular, the RNN block consists of $2$-layered GRUs \cite{GRU} with $128$ hidden units each.  The hidden units are all initialized with zero vectors in the simulation.  The GRU layers are followed by a fully connected layer that acts as the classifier.  In this work, the optimal future beam prediction problem is posed as a classification task.  For this, the softmax activation function is applied to the output of the final fully connected layer.  The recurrent architecture receives an input sequence of $r$ samples and predict the future beam indices $\hat{\mathbf{f}}_u[t_2]$ where $t_2 \in \{\tau + 1, \ldots, \tau + r^\prime\}$.

	\subsubsection{Proposed Position-Aided Solution} The baseline solution as presented in Section~\ref{sec:baseline_beam_track_sol} 
	utilizes the previously observed beam sequences to predict future beams.  One of the major drawbacks of this approach is the requirement for beam training.  In order to know the last beam, the base station needs to perform beam training.  The training overhead associated with the traditional beam training method is one of the major issues that we address in the paper.  In order to reduce the overhead, we propose an alternative and novel method to perform beam tracking using sensory data.  The real-time GPS position of the transmitter is one of the sensory data we study in this paper.  There are two main stages: (i) Data Processing and (ii) Recurrent Prediction.  The details of the two stages in the position-aided beam tracking solution are presented below:
	
	\textbf{Data Processing:} 
	The GPS measurements result in a position sample, which is two-dimensional, composed of geographical coordinates in decimal degrees, i.e., a latitude in $[-90^{\circ},90^{\circ}]$ and a longitude in $[-180^{\circ},180^{\circ}]$.  The positional data is normalized by applying a min-max normalization for both latitude and longitude values.  We first compute the minimum and maximum values across the dataset to perform the min-max normalization.  The values are then utilized to transform each data sample following $x' = (x - \min) / (\max - \min)$.  The normalized data is then provided as input to the next stage, i.e., recurrent prediction. 
	
	\textbf{Recurrent Prediction:} Similar to the baseline approach, the position-aided beam tracking solution adopts an RNN-based architecture to predict future beams.  The only difference with the baseline approach is that there is no embedding block required.  The GPS positions are normalized to lie within $[0, 1]$ as mentioned above.  The normalization enables faster and more stable training of the RNNs.  A $2$-layered GRU with $128$ hidden units in each layer is used for this task.  The output of the last unit in the second GRU is fed to a classifier to predict the future beams.  It is important to note here that the architecture of the recurrent layer is kept the same as the baseline approach to enable a fair comparison.

	\subsubsection{Proposed Vision-Aided Solution} The third modality of choice for the beam tracking task is the sequence of RGB images captured by the base station.  The successful prediction of the future beams using a sequence of images largely depends on the following two factors.  The ability to detect and identify relevant objects in the image is the first step in achieving the target.  In this study, since we have only one transmitter (drone) in the scene, the first step involves detecting and extracting relevant features from the images.  The first objective is attained using a state-of-the-art objection detection model.  Once we have the extracted feature vectors from all the images in a sequence, the second step is understanding the underlying relationship between each consecutive feature vector to predict the future beam indices.  Similar to the beam-only and position-aided beam tracking task, the second component involves an RNN-based model that takes the extracted feature vectors and predicts the future beams.  Next, we present in detail the two components of the proposed vision-aided beam-tracking solution.  
	
	\textbf{Feature Extraction:} The feature extraction phase aims to derive meaningful spatial representations from sequential image data for subsequent use in beam prediction. This work implements a modified version of YOLOv3 \cite{yolov3}, an advanced object detection framework built upon the Darknet-53 architecture, chosen for its demonstrated efficiency and accuracy in real-time object detection tasks \cite{Yolo}. The integration of YOLOv3 into the proposed framework necessitates two principal modifications. First, the network architecture is adapted specifically for drone detection by modifying the classifier layer to accommodate a single object class. This targeted modification preserves the pre-trained feature extraction capabilities while optimizing the network for drone-specific detection. Second, the modified YOLOv3 network undergoes fine-tuning using a domain-specific dataset that reflects the target wireless environment. This process optimizes both the modified classifier and the feature extraction layers to enhance detection performance for drone-based applications. For each input image, the optimized YOLOv3 model generates bounding box information for the detected drone. Specifically, the model outputs a $2 \times 1$ vector containing the normalized center coordinates $[x_{center}, y_{center}]$ of the bounding box, where coordinates are scaled to the range $[0,1]$. These normalized center coordinates constitute the feature vector that serves as input for the subsequent recurrent prediction phase. This dimensional reduction from complete bounding box information to center coordinates provides a compact yet informative spatial representation of the drone's position within the visual field.
	
	\textbf{Recurrent Prediction:} Similar to the beam-only and the position-aided beam tracking solution, the second stage of the proposed solution adopts an RNN-based model for predicting future beams.  The input to the model is the sequences of center coordinates extracted by the object detection model, as discussed above.  For a fair comparison, the recurrent architecture is kept the same for all three approaches. 
	
	%%%%%%%%%%%%%%%%%%%%%%%%%%%%%%%%%%%%%%%%%%%%%%%%
	
	%%%%%%%%%%%%%%%%%%%%%%%%%%%%%%%%%%%%%%%%%%%%%%%%
	\section{DeepSense Testbed and Dataset Analysis} \label{sec:testbed_and_dataset}
	
	Studying the value of sensing-aided solutions and deep learning in overcoming the beam training overhead challenge in mmWave/THz drone communication mandates the construction of a measurement-based real-world dataset. We modify the existing DeepSense 6G testbed to collect this dataset and build a new testbed comprising mmWave/THz drones. The resultant dataset introduced a new scenario (Scenario $23$) to the DeepSense 6G dataset. In the following sections, we will discuss the following. \textit{Scenario and Testbed Description} section describes the data collection location and the adopted DeepSense 6G testbed. \textit{Dataset Analysis} It provides a short analysis of the collected dataset. \textit{Task Specific Dataset} In this work, we target two specific problem statements, i.e., current beam prediction and future beam tracking. This section describes the details of the development dataset utilized for each task and highlights the similarities and differences between them. 
	
	\subsection{Scenario Description}\label{sec:scenario}
	
	%####################################################################################################
	\begin{figure}[t]
		\centering
		\subfigure[]{\centering \includegraphics[width=0.45\linewidth]{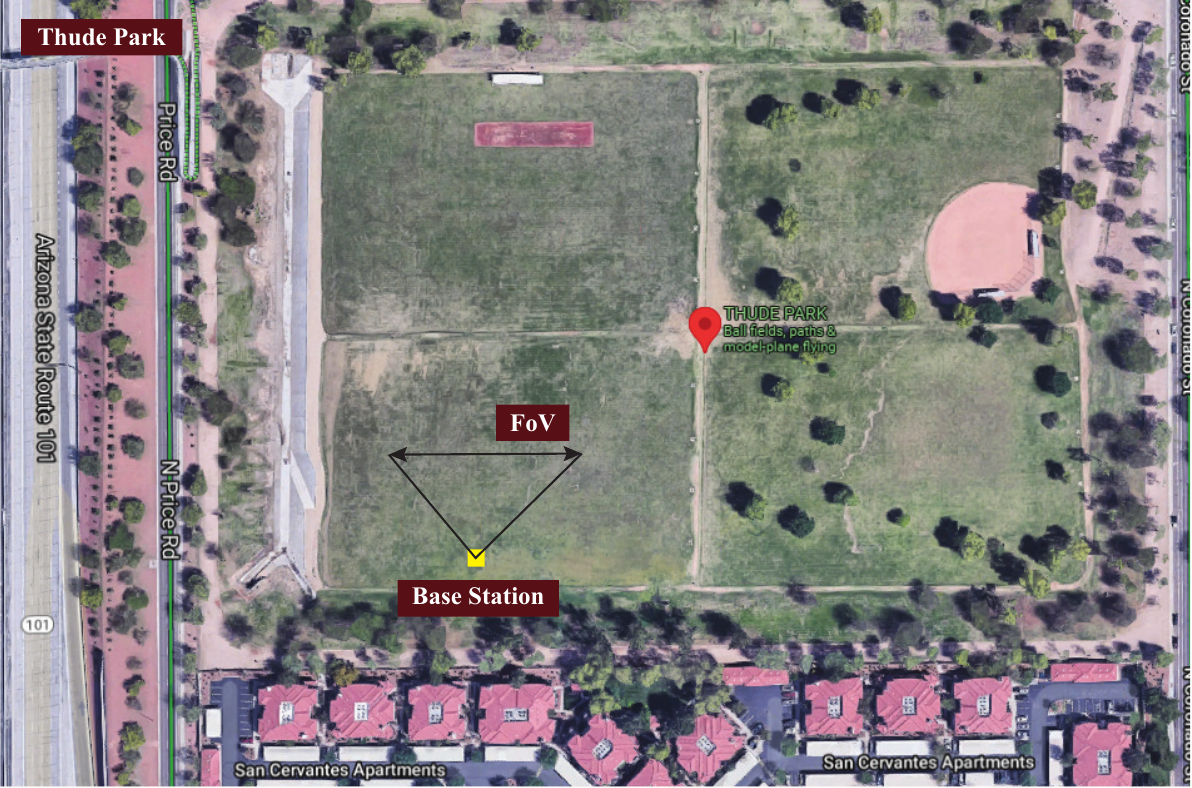}\label{fig:thude_park_map}}
		\subfigure[]{\centering \includegraphics[width=0.45\linewidth]{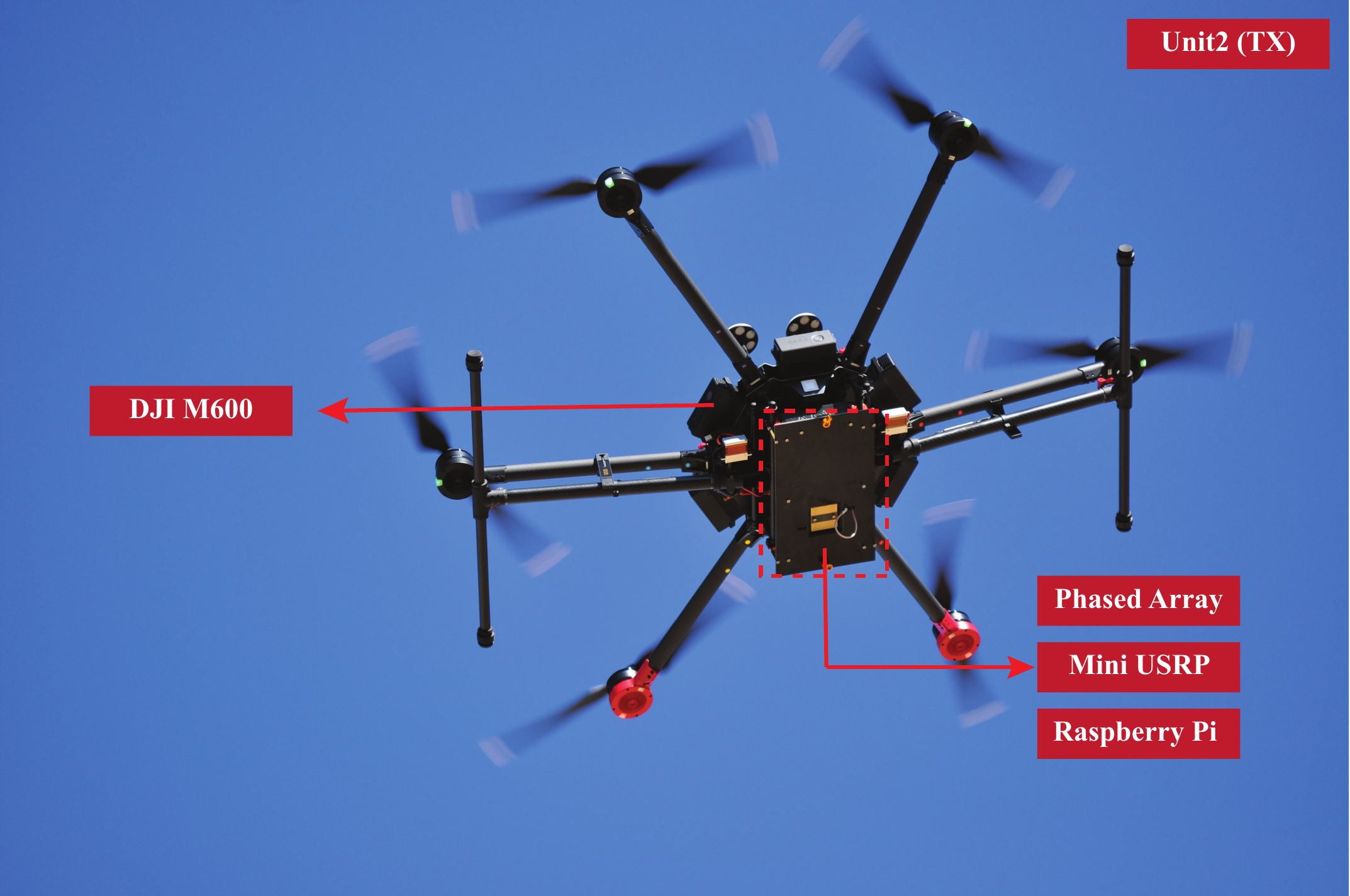}\label{fig:drone_view1}}
		\subfigure[]{\centering \includegraphics[width=0.45\linewidth]{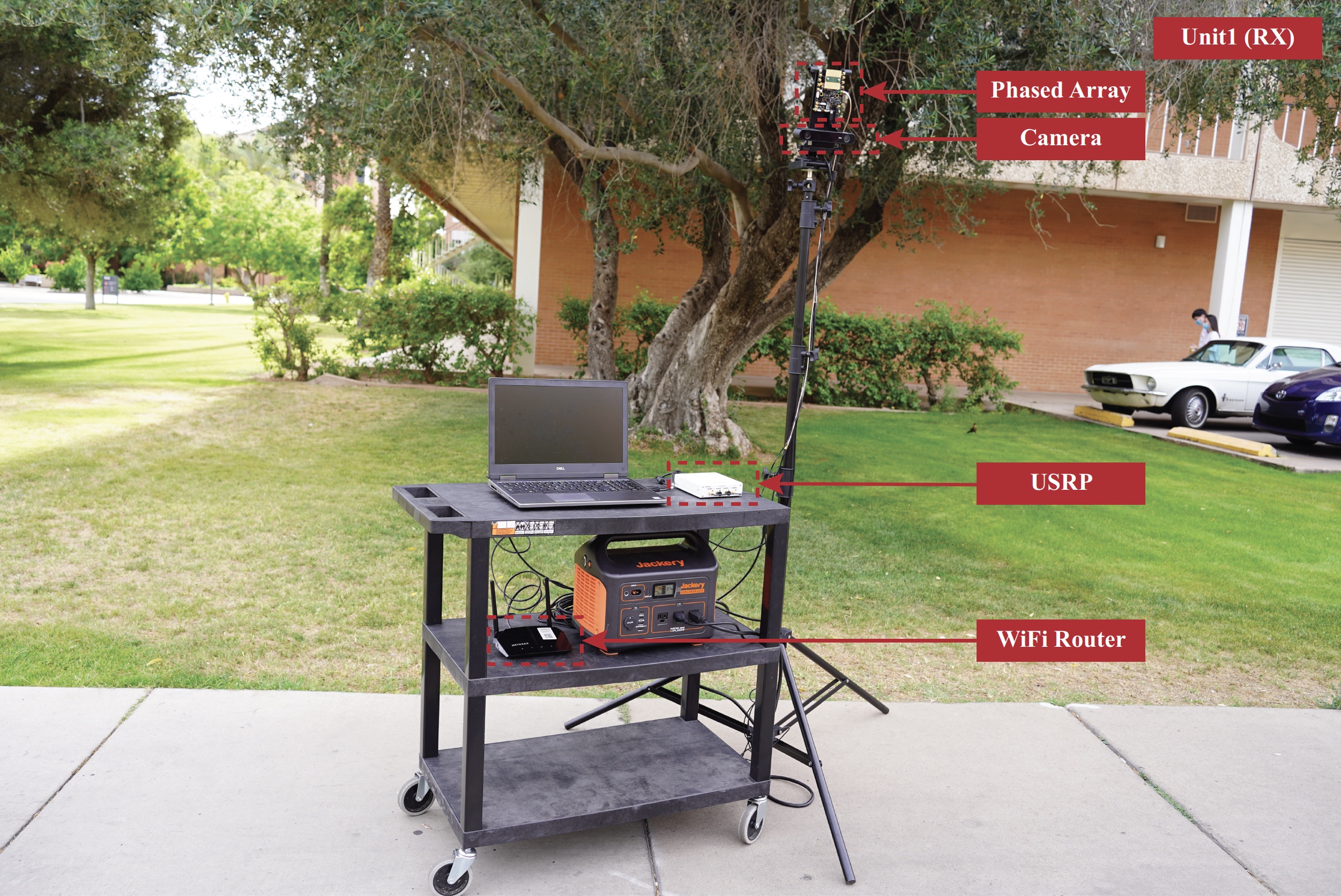}\label{fig:basestaion}}
		\caption{The figure illustrates the experimental setup and location of the DeepSense 6G testbed deployment. Fig. (a) presents an aerial perspective from Google Maps of Thude Park, the selected site for data collection. Fig. (b) and (c) detail the hardware configuration of both communication endpoints: the mobile drone-based transmitter and the stationary base station. Fig. (b) specifically depicts the drone-mounted mmWave phased array operating in the 60 GHz band that establishes the communication link with the base station.}
		\label{fig:scenario10_data_collection}
		%\vspace{-4mm}
	\end{figure}
	%####################################################################################################

	The experimental evaluation was conducted at Thude Park, a public rectangular park in Chandler, Arizona, with a total circumference of approximately 1.24 kilometers. The park features a dedicated model aircraft flying zone measuring 205 meters in length and 152 meters in breadth, providing an ideal and secure environment for drone-based experimentation. These dimensions enable comprehensive data collection across various distances, heights, and velocities relative to the base station, ensuring dataset diversity and real-world applicability. Fig.~\ref{fig:thude_park_map} illustrates the park's aerial view from Google Maps, indicating the base station's location, while Fig.~\ref{fig:drone_view1} demonstrates the data collection process with the mmWave phased array mounted beneath the drone.
	
	The experimental testbed was established at the park's Southwest corner (Fig.~\ref{fig:thude_park_map}). The stationary unit, designated as unit1 (RX), integrates a standard-resolution RGB camera with a 16-element ($M = 16$) mmWave phased array operating in the 60 GHz band. The receiver implements an over-sampled codebook comprising 64 pre-defined beams ($Q = 64$). The measurement equipment is positioned on an elevated platform approximately 1.5 meters above ground level, with both the camera and phased array oriented skyward to maximize the base station's field-of-view (FoV). The mobile unit, designated as unit2 (TX), consists of a remote-controlled drone equipped with multiple sensing and communication components: a mmWave transmitter featuring a quasi-omni antenna operating in the 60 GHz band with omnidirectional transmission capabilities, a GPS receiver, and inertial measurement units (IMU). To enhance dataset diversity, the experimental protocol incorporated various flight trajectories encompassing different altitudes, distances from the base station, and flight velocities. Detailed specifications of the experimental setup and data collection methodology are documented in \cite{DeepSense}.

	\begin{table}[!t]
		\caption{Scenario 18: Thude Park, Arizona}
		\label{table}
		\centering
		\setlength{\tabcolsep}{5pt}
		\renewcommand{\arraystretch}{1.4}
		\begin{tabularx}{\columnwidth}{|X|X|}
			\hline\hline
			\textbf{Testbed}             & 4                            \\ \hline
			\textbf{Number of Instances} & 12004                         \\ \hline
			\textbf{Number of Units}     & 2 \\ \hline
			\textbf{Total Data Modalities}     & RGB images, mmWave received power, GPS positions \\ \hline\hline
			\multicolumn{2}{|c|}{\textbf{Unit 1}} \\ \hline
			\textbf{Type} & Stationary \\ \hline
			\textbf{Hardware elements} & RGB Camera, mmWave phased array \\ \hline
			\textbf{Data Modalities} & RGB images and 64-mmWave received power \\
			\hline\hline
			\multicolumn{2}{|c|}{\textbf{Unit 2}} \\ \hline
			\textbf{Type} & Mobile \\ \hline
			\textbf{Hardware elements} & GPS receiver, IMU, and mmWave phased array \\ \hline
			\textbf{Data Modalities} & GPS positions, Height, Compass heading, Pitch and Yaw  \\ 
			%	\textbf{Collected Data Modalities} & GPS position \\
			\hline\hline
		\end{tabularx}
		\label{ref:tab_scenario_18}
	\end{table}

	\subsection{Dataset Analysis}\label{sec:dev_data_analysis} 
	
	\begin{figure}[t]
		\centering
		\subfigure[]{\centering \includegraphics[width=0.32\linewidth]{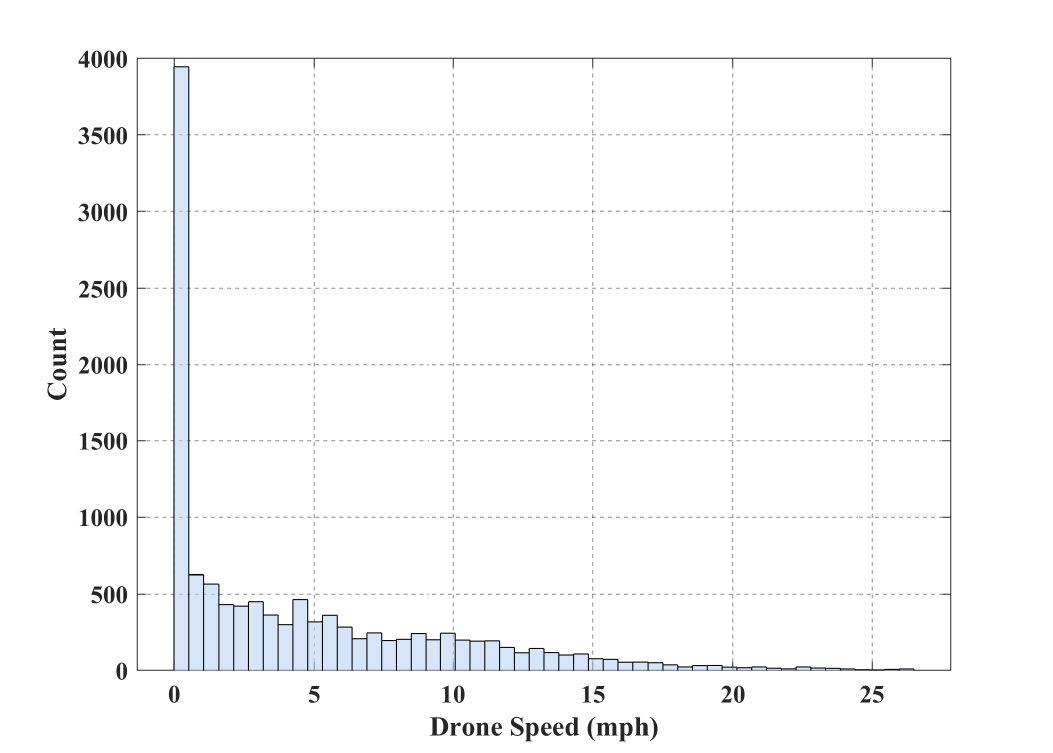}\label{fig:speed_dist}}
		\subfigure[]{\centering \includegraphics[width=0.32\linewidth]{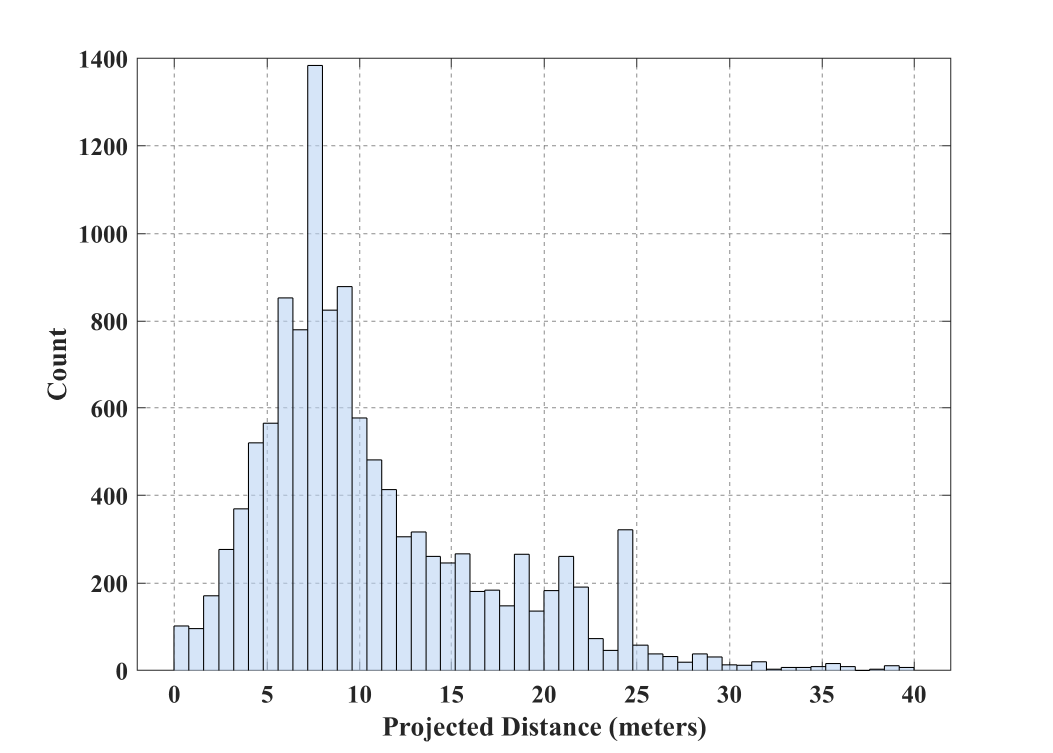}\label{fig:dist_hist}}	
		\subfigure[]{\centering \includegraphics[width=0.32\linewidth]{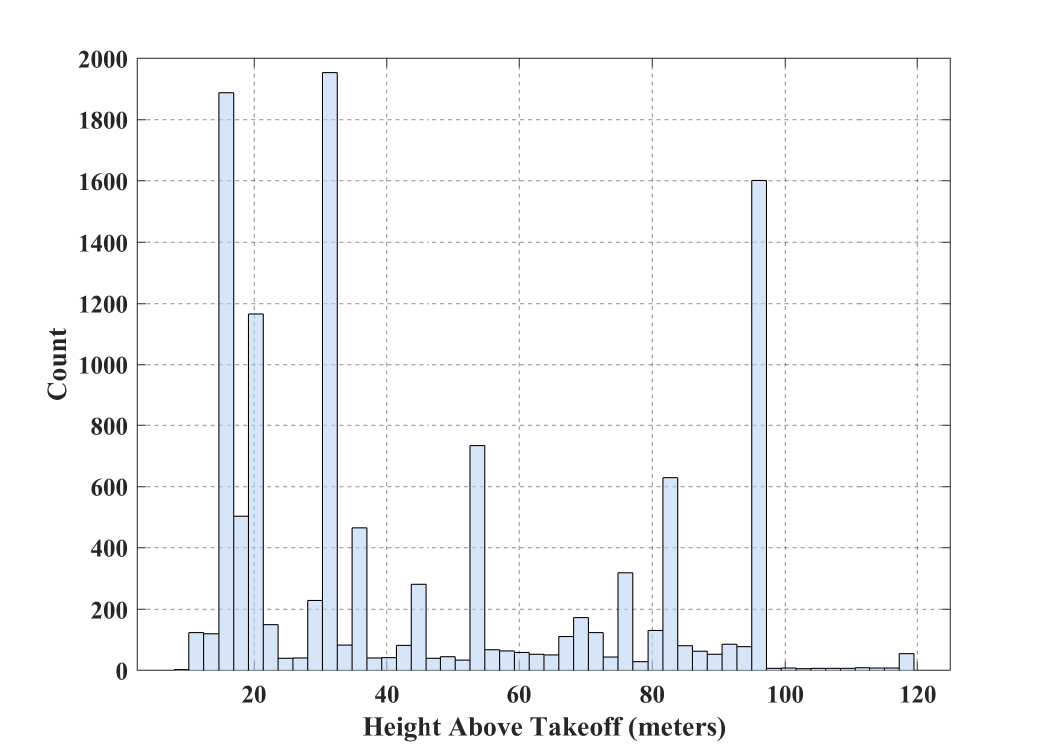}\label{fig:height_hist}}		
		\subfigure[]{\centering \includegraphics[width=0.32\linewidth]{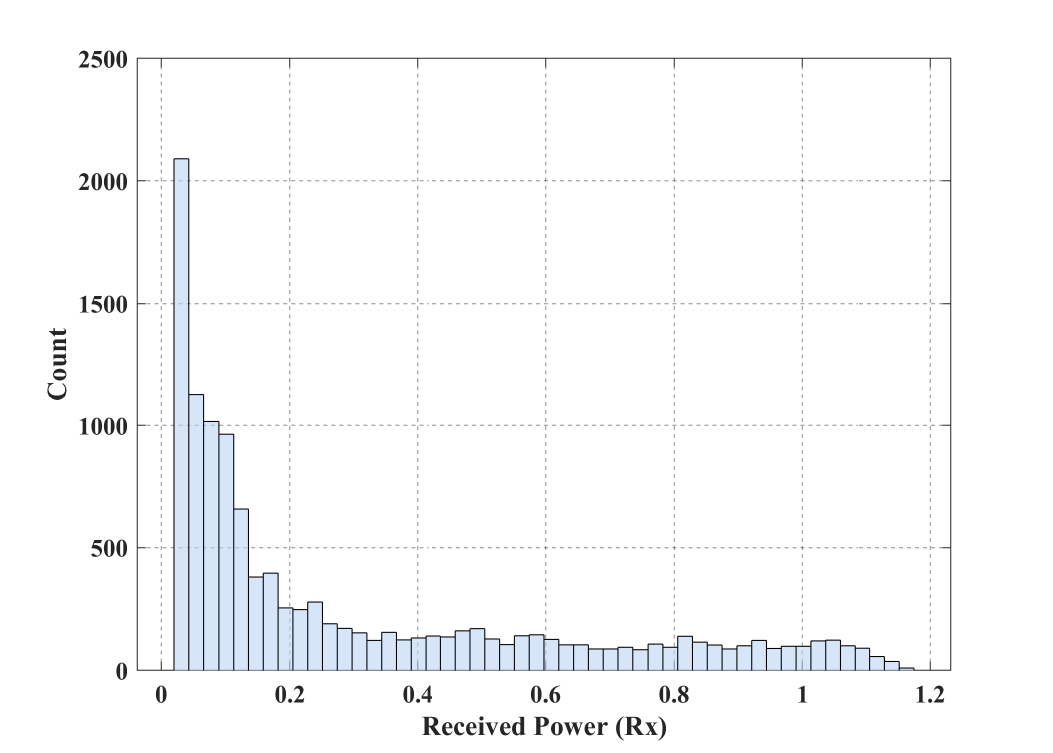}\label{fig:power_dist}}	
		\subfigure[]{\centering \includegraphics[width=0.32\linewidth]{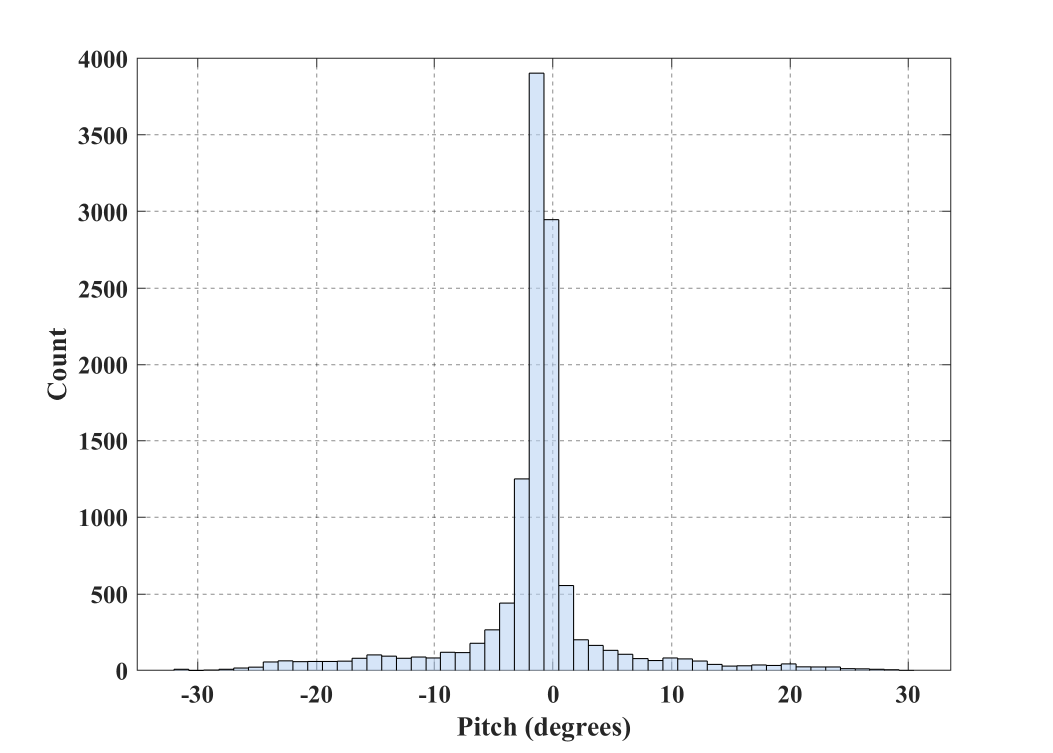}\label{fig:pitch_hist}}	
		\subfigure[]{\centering \includegraphics[width=0.32\linewidth]{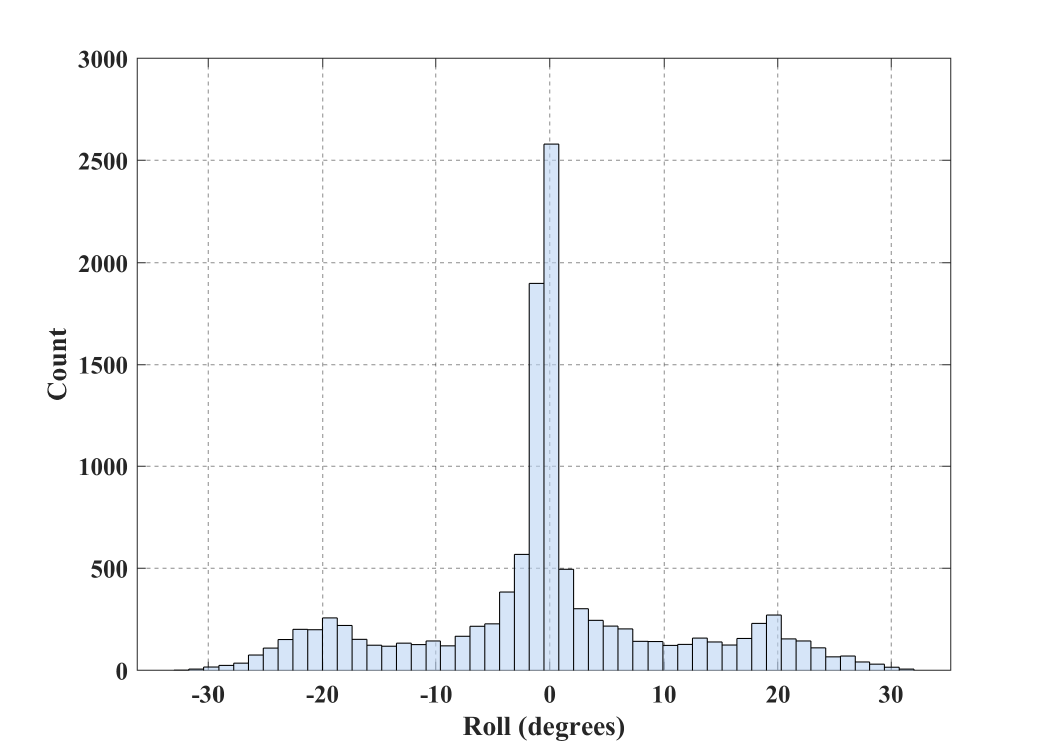}\label{fig:roll_hist}}		
		\caption{Figures (a), (b) and (c) shows the distribution of speed, distance and height of the transmitting drone in this dataset. In figure (d), we present the variance of the received power across the dataset. Figures (e) and (f), present the distribution of pitch and roll movements, respectively in the dataset.}
	\end{figure}

	DeepSense 6G testbed v4 is utilized to collect the multi-modal dataset targeting 6G drone communication. This particular dataset forms Scenario 23 of the DeepSense 6G database \cite{DeepSense}. The previous subsection presented an overview of the location used for this data collection. In this subsection, we provide an in-depth analysis of the dataset. The dataset comprises different modalities such as the wireless received power, images, GPS position of the transmitting drone, height, distance, and speed of the drone, pitch, and roll information. In order to quantify the diversity, first, we present the distribution of different data modalities. Then, we study the relationship between the wireless beam index and other data modalities, such as the position of the drone and the height of flight. Lastly, we study the dependence of the maximum received power on the drone's speed and the pitch angle.

	\subsubsection{Dataset Distribution}
	DeepSense 6G is envisioned and built to be a multi-modal and diverse real-world database. Unlike the other scenarios in the database with humans or vehicles as the transmitting candidate, this particular scenario (23) utilizes drones as the transmitter. This adds diversity to the overall database and provides an interesting opportunity to study the different dynamics of 6G drone communication. The dataset comprises different data modalities as presented earlier in Section~\ref{sec:scenario}. It is important to capture and highlight the variability in different modalities in the dataset. To that end, we start the analysis by first presenting the distributions of different data modalities. The drone's variability in speed, distance, and height is an important aspect of this dataset. It captures the extent of the effective wireless environment and the movement patterns of the transmitter. The height and distance of the transmitter have a direct impact on the visual data. For example, as observed from the stationary unit, the aspect ratio of the transmitter will vary as the height and distance change. The varying scale of the objects in the image poses a significant challenge and is an important factor in testing the practicality of any sensing-aided machine learning algorithm. The varying speed of the drone also poses a challenge for different wireless communication tasks, such as current beam prediction, future beam tracking, and blockage prediction in high-mobility communication settings. Fig.~\ref{fig:speed_dist}, ~\ref{fig:dist_hist}, and \ref{fig:height_hist} show the distribution of speed, distance, and height of the transmitting drone in this dataset. Fig.~\ref{fig:speed_dist} shows the drone traveling at different speeds up to $\approx 25$ miles per hour. There is a peak at zero speed, which refers to the drone hovering at the same spot. It is important to note here that this natural variance in speed makes the dataset versatile and different from a vehicular or human dataset. Another important feature specific to the drone dataset is the variance in the height and distance of the transmitter. The drone is flown at varying heights and distances from the base station as shown in Fig.~\ref{fig:dist_hist} and \ref{fig:speed_dist}. The variability of mmWave received power distribution is another factor that could help highlight the dataset's diversity. The received power is a function of the transmitter's distance, height, and orientation with respect to the stationary unit. This dependence can be observed in the variance of the received power, as shown in Fig.~\ref{fig:power_dist}. Unlike land-bound vehicles, drones have the flexibility of moving in three dimensions. Along with moving forward, backward, and sideward, the drone can gain or lose altitude. It can also change orientation through rotation along the three axes to produce yaw (normal axis), pitch (transverse axis), and roll (longitudinal axis). Fig.~\ref{fig:pitch_hist} and \ref{fig:roll_hist} present the distribution of pitch and roll movements, respectively, in the dataset.

	\begin{figure}[t]
		\centering
		\subfigure[Speed]{\centering \includegraphics[width=0.32\linewidth]{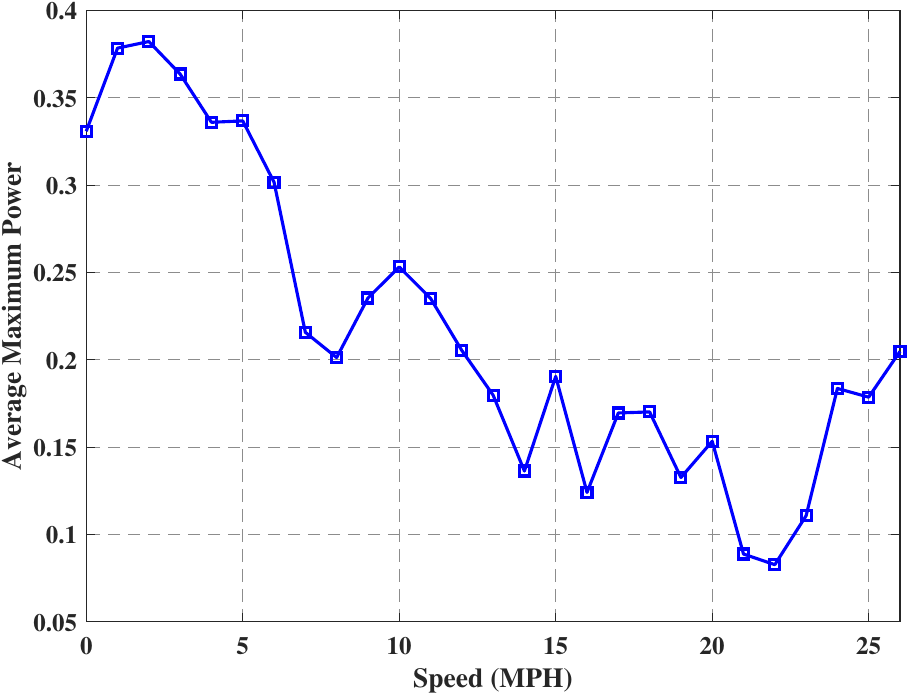}\label{fig:max_speed_vs_power}}	
		\subfigure[Pitch]{\centering \includegraphics[width=0.365\linewidth]{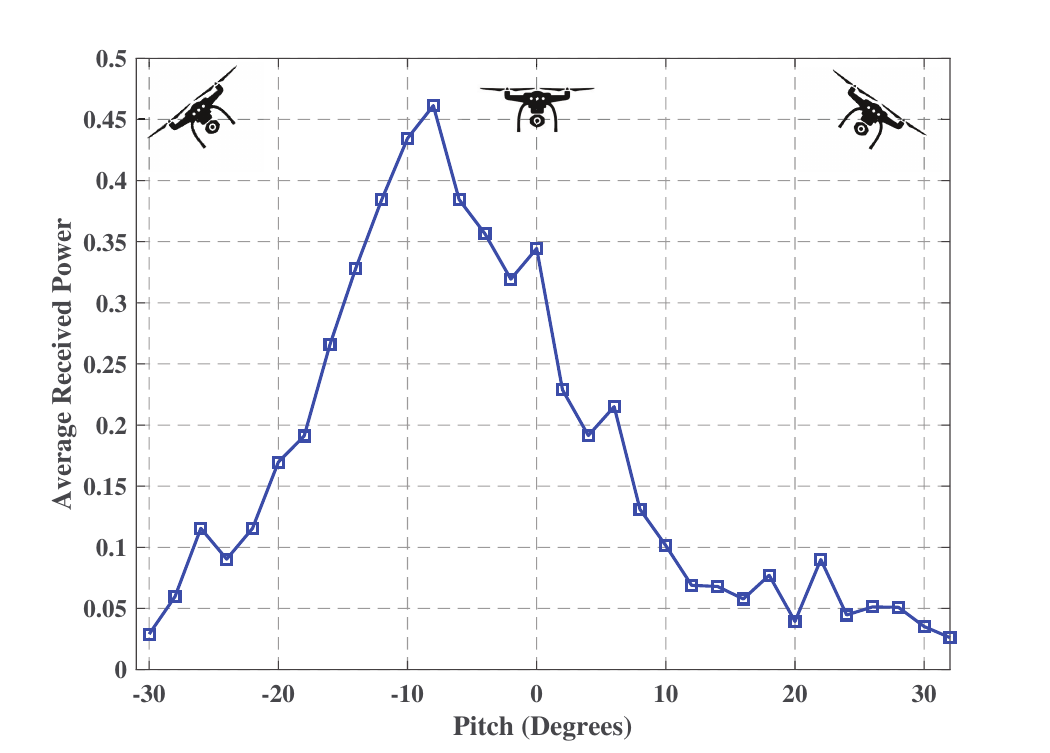}\label{fig:pitch_vs_power}}		
		\subfigure[Distance]{\centering \includegraphics[width=0.32\linewidth]{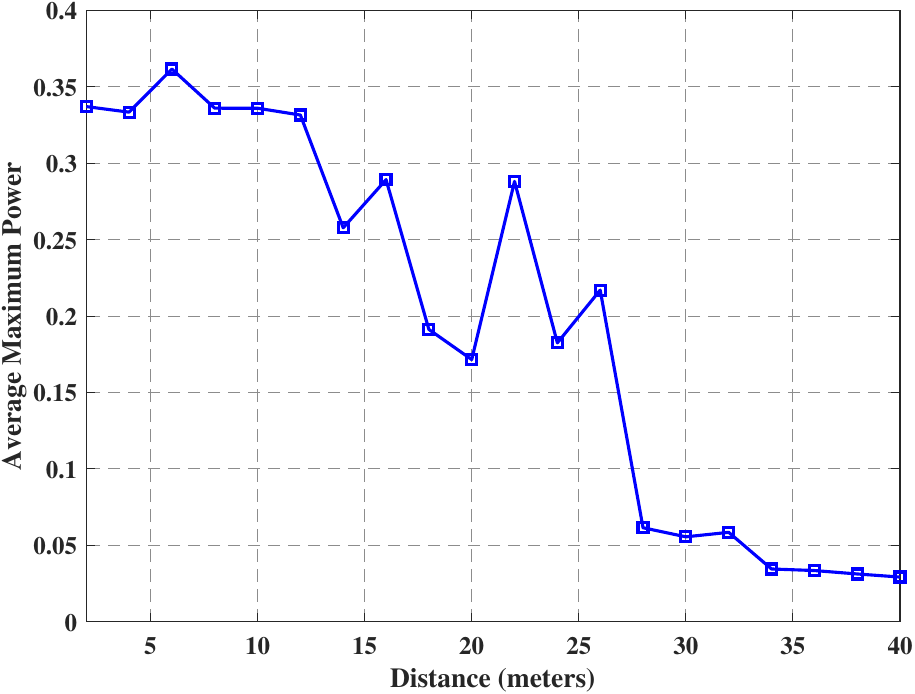}\label{fig:max_pwr_vs_distance}}	
		\caption{Figures (a), (b), and (c) show the distribution of the maximum receive power as the function of speed, pitch, and distance, respectively.  }
		\label{fig:max_pwr_dist}
	\end{figure}

	\subsubsection{Maximum Received Power Distribution}
	
	DeepSense 6G is envisioned as a multi-modal dataset to enable joint wireless communication and sensing research. One of the most critical modalities in this dataset is the mmWave received power. The receive power is a function of environmental factors such as the location and the presence or absence of other objects in the environment, to name a few; it captures the wireless channel behavior in a real wireless setting. The variance in the received power across the different samples highlights the diversity in the dataset. In Fig.~\ref{fig:max_pwr_dist}, we show the dependence of the average received power on the different features of the dataset, such as speed of travel, pitch, and the distance of the drone from the stationary base station. In Fig.~\ref{fig:max_speed_vs_power} and \figref{fig:max_pwr_vs_distance}, we observe that there is a correlation between the speed and distance of the drone with the average maximum received power. With the increase in both the distance and the speed of the drone, the average received power decreases. In general, the decrease is expected for the distance plot as the received power follows an inverse square law, i.e., it is proportional to the inverse square of the distance. However, the impact of speed on maximum power is an interesting observation. It follows a similar trend to Fig.~\ref{fig:max_pwr_vs_distance}, i.e., the average received power decreases as the speed increases. The decrease can probably be attributed to the physics behind the drone motion. Instead of staying horizontal to the ground while moving forward, the drones need to tilt downward, explaining the decreased receive power as the speed increases. Another interesting observation is shown in Fig.~\ref{fig:pitch_vs_power}, which plots the average received power versus the pitch. Pitch is the rotation of the drone along the lateral or transverse axis. It is a measure of the tilt upwards or downwards based on the orientation of the drone. As mentioned in Section~\ref{sec:scenario}, the phased array is mounted on the bottom of the drone facing downwards. As the pitch changes, i.e., as the drone tilts upwards or downwards, the orientation of the phased array mounted on the drone also varies. We observe a peak in the average received power when the tilt of the drone is relatively lower.

	\subsection{Task Specific Dataset}\label{sec:task_specific_data} 
	
	This work focuses on two crucial mmWave drone communication tasks: (i) sensing-aided beam prediction task and (ii) sensing-aided beam tracking task. The two communication task and their proposed solutions mandate different types of datasets. This is due to the inherent difference between the two tasks; the beam prediction task focuses on predicting the current optimal beam, whereas the beam tracking task predicts the future optimal beams. As mentioned in Section~\ref{sec:dev_data_analysis}, the dataset undergoes a processing pipeline that results in two different development datasets.

	\begin{table*}[!t]
		\caption{Sensing-Aided Beam Prediction and Tracking Dataset}
		\centering
		\setlength{\tabcolsep}{5pt}
		\renewcommand{\arraystretch}{1.4}
		\begin{tabular}{|c|c|c|c|}
			\hline
			\multirow{2}{*}{\textbf{Task}}                          & \multirow{2}{*}{\textbf{Sensing Modalities}} & \multirow{2}{*}{\textbf{Dataset}} & \textbf{Number of Samples} \\ \cline{4-4} 
			&                                              &                                   & \textbf{Training}          \\ \hline
			\multirow{3}{*}{\textbf{Sensing-Aided Beam Prediction}} & GPS Positions                                & $ \mathcal D_{\text{task}_{1-1}}$ & 12005                      \\ \cline{2-4} 
			& GPS Positions + Height + Distance            & $ \mathcal D_{\text{task}_{1-2}}$ & 12005                      \\ \cline{2-4} 
			& RGB Image                                    & $ \mathcal D_{\text{task}_{1-3}}$ & 12005                      \\ \hline
			\multirow{3}{*}{\textbf{Sensing-Aided Beam Tracking}}   & Beam-only                                    & $\mathcal D_{\text{task}_{2-1}}$  & 11985                      \\ \cline{2-4} 
			& GPS Positions                                & $\mathcal D_{\text{task}_{2-2}}$  & 11985                      \\ \cline{2-4} 
			& RGB Image                                    & $\mathcal D_{\text{task}_{2-3}}$  & 11985                      \\ \hline
		\end{tabular}
		\label{ref:tab_dataset_analysis}
		%\vspace{-4mm}
	\end{table*}

	\textbf{Sensing-Aided Beam Prediction Dataset:} The beam prediction task, as defined in Section~\ref{sec:beam_pred}, focuses on predicting current optimal beam indices using available sensing information. This task requires evaluation using data collected from a real wireless environment with a drone serving as the mmWave transmitter. The details of the DeepSense 6G testbed, the data collection location, and the comprehensive data analysis are presented in Section~\ref{sec:testbed_and_dataset} - \ref{sec:scenario} and Section~\ref{sec:testbed_and_dataset} - \ref{sec:dev_data_analysis}, respectively. The collected dataset encompasses multiple data modalities: RGB images captured at the base station, real-time GPS location information from the drone, spatial parameters (distance, height, speed, and orientation) of the drone, and a $64 \times 1$ vector representing mmWave received power measurements. The dataset preparation involves a three-step processing pipeline. The initial step consists of filtering data samples where the drone operates outside the visual field-of-view (FoV) of the base station, accomplished through manual examination of all collected samples. The second step implements downsampling of the $64 \times 1$ power vector to a $32 \times 1$ dimension ($Q = 32$) by selecting alternate samples from the vector. This downsampling operation preserves the total coverage area of the beams, as the base station employs an oversampled codebook of $64$ pre-defined beams. For each data sample, the beam index corresponding to the maximum received power in the downsampled power vector is designated as the optimal beam value index. The final processing step involves partitioning the dataset into training and test sets following a 70-30\% distribution. Table~\ref{ref:tab_dataset_analysis} presents the detailed characteristics of the development datasets $[\mathcal D_{\text{task}_{1-1}}, \mathcal D_{\text{task}_{1-2}}, \mathcal D_{\text{task}_{1-3}}]$ utilized across the three sub-tasks in the mmWave drone beam prediction task. While the total number of samples remains consistent across all three beam prediction sub-tasks, the distinction lies in the specific data modalities employed for beam prediction in each case.

	\textbf{Sensing-Aided Beam Tracking Dataset:} Similar to the beam prediction task, the mmWave drone beam tracking task requires samples from the real wireless environment. Instead of observing a data sample and predicting the optimal beam, the beam tracking task involves observing a sequence of $8$ data samples ($r = 8$) and predicting the future beams. This work focuses on predicting up to three future beams, i.e., $r^\prime = 3$. In order to generate the final development dataset for the beam tracking task, we adopt a similar processing pipeline as described above. The first two steps, involving data filtering and power vector downsampling, remain the same. For this task, the third step involves formatting the dataset such that each sample now is a tuple consisting of a sequence of $8$ data samples ( optimal beam index, images, the GPS positions) and optimal beam indices for the three future time steps. The final dataset is then further divided into training and test sets following a similar $70-30\%$ split. Table~\ref{ref:tab_dataset_analysis} also shows the details of the dataset used for the three sub-tasks as defined in ~\ref{sec:beam_track}.

	The two development datasets above are then utilized for training and evaluating the performance of the proposed solutions. We provide the details of the training in the next section while Section~\ref{sec:beam_pred_eval} and Section~\ref{sec:beam_track_eval} discuss in detail the performance of the proposed beam prediction and beam tracking solutions, respectively.

	\section{Beam Prediction Evaluation}\label{sec:beam_pred_eval}
	This section studies the performance of the proposed solutions for the sensing-aided beam prediction task. In the first sub-section, we will present the details of the experimental setup, which consists of the neural network training parameters and the adopted evaluation metric for the beam prediction task. Next, we will discuss the performance of the proposed solution for each of the three sub-tasks. In general, we perform a comparative study of the different sensing modalities utilized for this beam prediction task. It highlights and discusses the main advantages and drawbacks of each modality centered around the beam prediction task.
	
	\subsection{Experimental Setup}\label{sec:beam_pred_setup}
	The proposed beam prediction framework incorporates three distinct sensing modalities: (i) position information alone, (ii) position combined with height and distance measurements, and (iii) visual data. Each modality necessitates a specific machine learning architecture as detailed in Section~\ref{sec:beam_pred_prop_soln}. This section presents the training hyperparameters for each machine learning model and establishes the evaluation criteria for assessing beam prediction performance.
	
	%######################################################################################
	%######################################################################################
	
	\begin{table*}[!t]
		\caption{Beam Prediction: Design and Training Hyper-parameters}
		\centering
		\setlength{\tabcolsep}{5pt}
		\renewcommand{\arraystretch}{1.2}
		\begin{tabular}{@{}l|ccc@{}}
			\toprule
			\toprule
			\textbf{Parameters}                     & \textbf{Vision}  & \textbf{Position}        & \textbf{Position-Distance-Height}       \\ \midrule \midrule
			\textbf{ML Model}                       & ResNet-50           & 2-layered MLP       & 2-layered MLP                 \\
			\textbf{Batch Size}                     & 32                  & 32                  & 32                 \\
			\textbf{Learning Rate}                  & $1 \times 10 ^{-4}$ & $1 \times 10 ^{-2}$ & $1 \times 10 ^{-2}$\\
			\textbf{Learning Rate Decay}            & epochs 4, 8 and 12       & epochs 20, 40 and 80    & epochs 20, 40 and 80   \\
			\textbf{Learning Rate Reduction Factor} & 0.1                 & 0.1                 & 0.1                \\
			%\textbf{Dropout}                        & 0.3                 & 0.3                 & 0.3                \\
			\textbf{Total Training Epochs}          & 20                  & 100                 & 100                \\ \bottomrule \bottomrule
		\end{tabular}
		\label{tab_beam_pred_train_params}
	\end{table*}

	%######################################################################################
	%######################################################################################
	
	\textbf{Network Training:} The implementation encompasses three distinct machine learning models corresponding to the different sensing modalities. For the position-based approaches—both position-alone and position with height and distance—the architecture employs 2-layer fully-connected neural networks (2-layered MLP). These models are trained with a batch size of 32 and an initial learning rate of $1 \times 10^{-2}$. The learning rate is adjusted at epochs 20, 40, and 80 with a reduction factor of 0.1, and the training continues for a total of 100 epochs. The vision-aided solution implements a ResNet-50 architecture with distinct training parameters: a batch size of 32, an initial learning rate of $1 \times 10^{-4}$, and learning rate adjustments at epochs 4, 8, and 12. The vision model training extends for 20 epochs. All models utilize cross-entropy loss with the Adam optimizer during the training phase, and the complete set of hyperparameters is detailed in Table~\ref{tab_beam_pred_train_params}.
	
	\textbf{Evaluation Criteria:} The performance assessment of the proposed sensing-aided beam prediction solutions employs two primary metrics. First, the top-k accuracy metric quantifies the percentage of test samples where the ground-truth optimal beam appears within the top-$k$ predicted beam indices. To ensure comprehensive evaluation, the analysis presents accuracy measurements across multiple k values: top-1, top-2, top-3, and top-5 accuracy. Additionally, the R2-score between the ground-truth top-1 power and the predicted power is computed to evaluate the quality of power prediction. This metric provides insight into how well the predicted beam's power correlates with the optimal beam's power, offering a complementary performance measure to the accuracy metrics.

	\subsection{Experimental Results} \label{sec:beam_pred_results}

	With the experimental setup as described in Section~\ref{sec:beam_pred_setup}, in this subsection, we study the beam prediction performance of the proposed solution. In general, we study the performance from two different standpoints: machine learning and wireless communication. In the first set of studies, we evaluate the performance of the proposed solutions from a machine learning perspective, i.e., beam prediction accuracy per approach, number of samples required for training, etc. The second set of studies looks into the performance from the wireless perspective, such as studying the implications of missed predictions on the wireless received power, etc. We started this paper by highlighting some of the important questions in Section~\ref{sec:Intro} that need to be answered to enable mmWave drone communication in a real wireless setting. In this subsection, we try to answer the questions pertaining to the beam prediction task. 
	
	\subsubsection{Can sensing-based beam prediction approaches for mmWave drone communication be an alternative to the traditional beam training method? }

\begin{figure}[!t]
		\centering
		\includegraphics[width=0.8\linewidth]{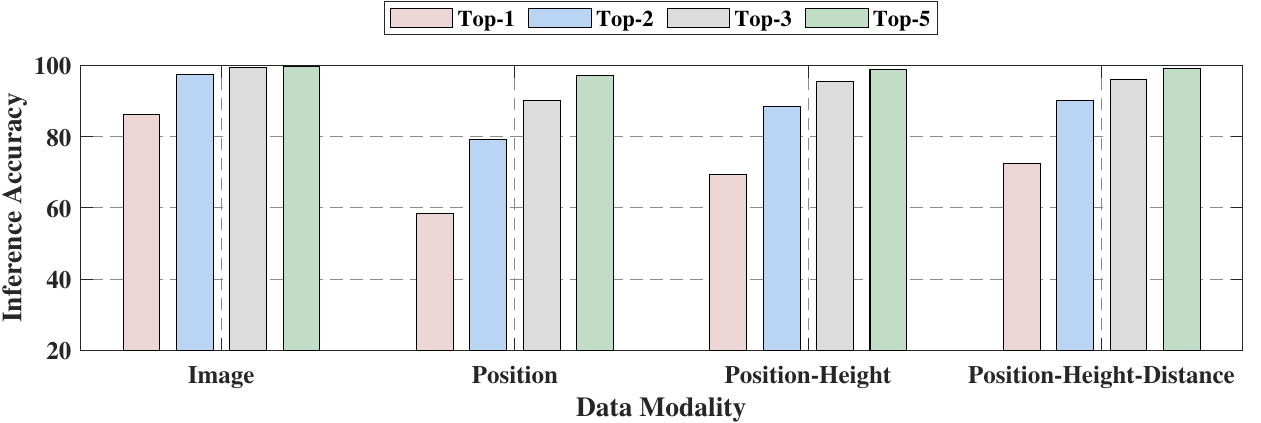}
		\caption{This figure plots the top-k accuracies $(k \in (1,2,3,5))$ for the proposed sensing-aided beam prediction solution. It is observed the vision-aided beam prediction solution outperforms the other approaches. Another interesting observation is that the performance improves as we incorporate additional sensing data with GPS positions. }
		\label{fig:beam_pred_acc}
	\end{figure}

	Prior investigations in sensing-aided beam prediction focusing on vehicular or human scenarios have demonstrated promising results \cite{charan2021c, Alrabeiah_camera, Morais22}. However, drone-based user equipment presents distinct challenges due to its six degrees of freedom in motion and variable orientation capabilities. This analysis evaluates whether sensing-aided solutions can maintain comparable beam prediction performance in drone-based transmission scenarios. The evaluation utilizes the drone beam prediction dataset described in Section~\ref{sec:task_specific_data}, with the traditional beam training approach serving as the baseline and upper performance bound. The experimental results presented in Fig.~\ref{fig:beam_pred_acc} reveal several significant findings. The position-only approach achieves a top-1 accuracy of approximately 59\%, indicating that positional information alone is insufficient for accurate beam prediction in drone-based mmWave communications. This limitation is further evidenced by the performance improvements observed when incorporating additional spatial parameters. The integration of height information and the combination of height and distance measurements yield accuracy improvements of approximately 10-14\% over the position-only solution. These quantitative improvements demonstrate the necessity of incorporating additional spatial parameters beyond mere positional data. The vision-based solution demonstrates superior performance among all evaluated approaches, achieving top-1, top-3, and top-5 accuracies of 86.32\%, 99.41\%, and 99.69\%, respectively. This enhanced performance can be attributed to the comprehensive spatial information captured in visual data. Images inherently encode multiple spatial attributes: the object's orientation, its position within the field of view, and its relative scale—which correlates with distance. This intrinsic ability to capture multiple spatial dimensions simultaneously enables the vision-based approach to outperform other methods. The performance differential underscores the significance of utilizing rich, multi-dimensional sensory data for accurate beam prediction in drone-based communications. These findings provide quantitative evidence supporting the necessity of comprehensive spatial information for accurate beam prediction in drone-based mmWave systems, where traditional position-only approaches prove insufficient due to the complex spatial dynamics involved.
	
	\subsubsection{How many data samples are needed to achieve the optimum prediction accuracy?}
	
\begin{figure}[t]
		\centering
		\subfigure[Vision]{\centering \includegraphics[width=0.45\linewidth]{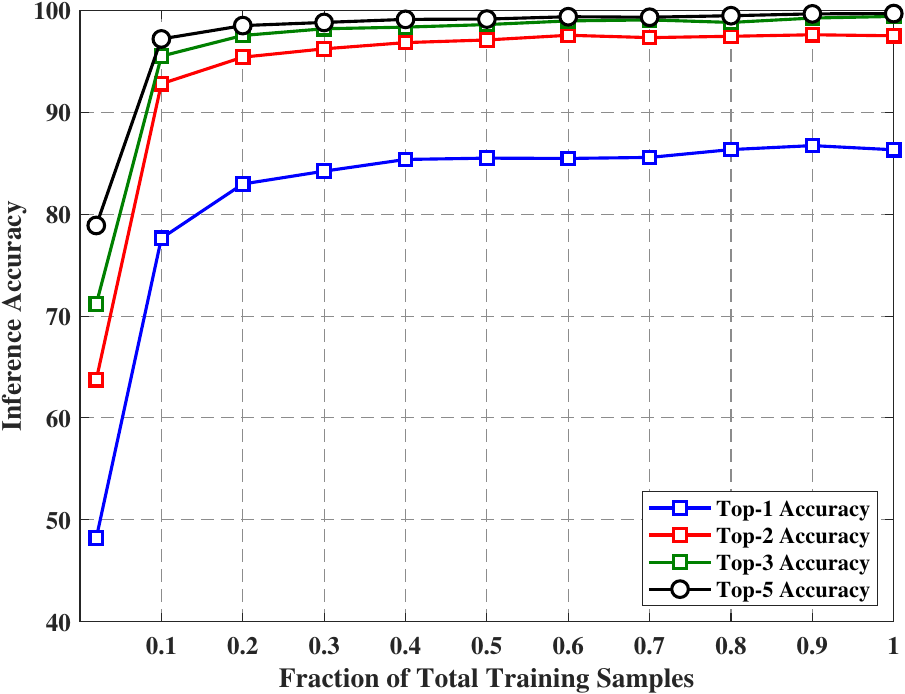}\label{fig:vision_num_samples}}
		\subfigure[Position]{\centering \includegraphics[width=0.45\linewidth]{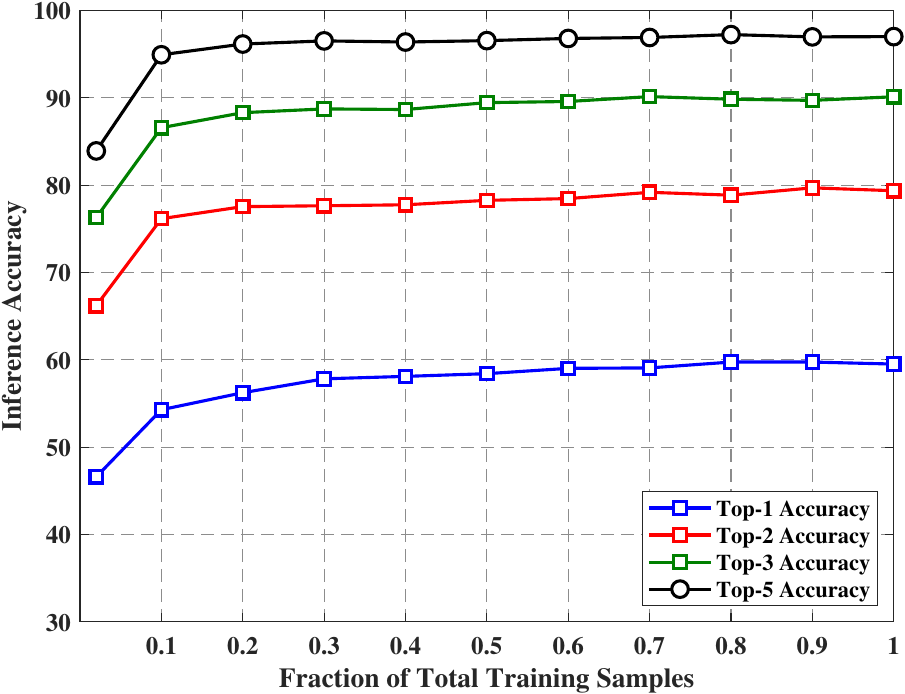}\label{fig:pos_num_samples}}
		
		\caption{This figure plots the top-1, top-2, top-3 and top-5 beam prediction accuracy versus the different dataset sizes for (a) vision-aided solution and (b) position-aided beam prediction solution. The figure highlights only $\approx 40\%$ of the training samples is needed to achieve the optimum beam prediction performance and that adding more training samples will not necessarily improve the performance of the proposed ML models.   }
		\label{fig:num_training_samples}
	\end{figure}

	The previous discussion related to the beam prediction performance of the proposed solutions leads us to this next discussion, where we study an interesting question regarding the number of data samples needed to achieve optimum performance. One of the major limitations of a deep learning-based solution is the need for a large amount of labeled data to achieve satisfactory performance. Therefore, one can argue that the lower performance of position-aided solutions is due to insufficient training samples. As presented in Section~\ref{sec:task_specific_data}, for the beam prediction task, we have a total of $8403$ training samples and $3602$ test samples. Fig.~\ref{fig:num_training_samples} provides an answer to this particular question of the number of samples needed. Fig.~\ref{fig:vision_num_samples} and Fig.~\ref{fig:pos_num_samples} plots the beam prediction accuracy versus the percentage of training samples required for vision-aided and position-aided beam prediction solutions, respectively. An obvious observation in both the figures is that the proposed solutions can learn the mmWave drone beam prediction task with $\approx 30 - 40\%$ of the training samples. The figures also highlight that adding more training samples to the dataset does not impact the performance of the ML model. Fig.~\ref{fig:pos_num_samples} consolidates the earlier conclusion that position alone is not sufficient for the mmWave drone beam prediction task, and the lower performance is indeed due to the lack of more information rather than the need for more training samples.

	\subsubsection{Do missed predictions act as roadblocks in sensing-aided beam prediction?}
	
	\begin{figure}[t]
		\centering
		\subfigure[Vision]{\centering \includegraphics[width=0.32\columnwidth]{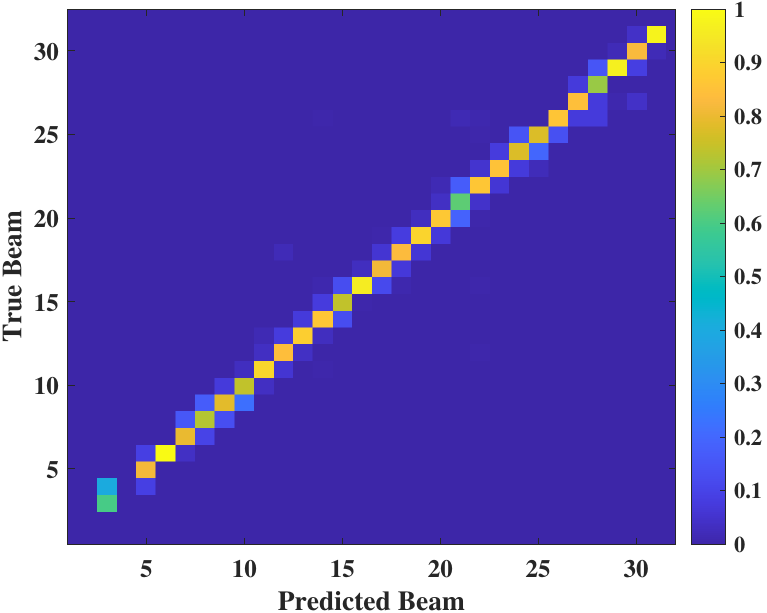}\label{fig:img_beam_pred_cf}}
		\subfigure[Position]{\centering \includegraphics[width=0.32\columnwidth]{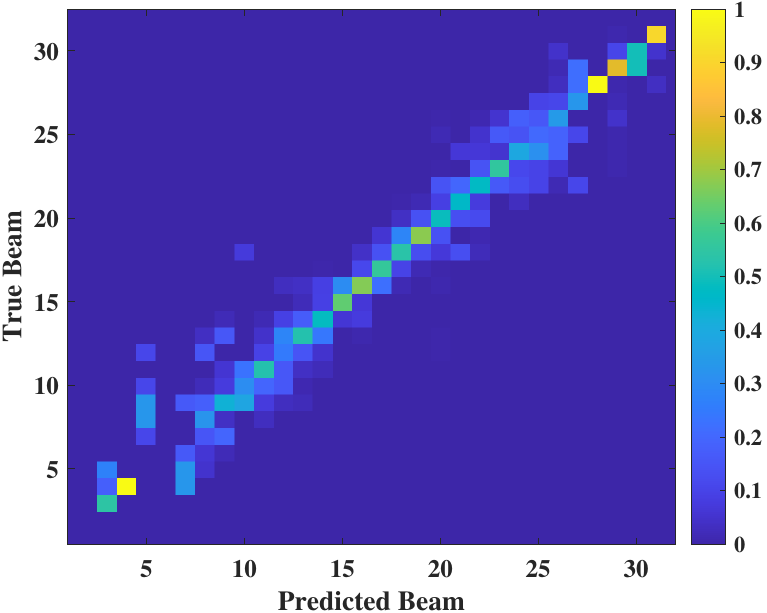}\label{fig:pos_beam_pred_cf}}
		\subfigure[Position, Height, and Distance]{\centering \includegraphics[width=0.32\columnwidth]{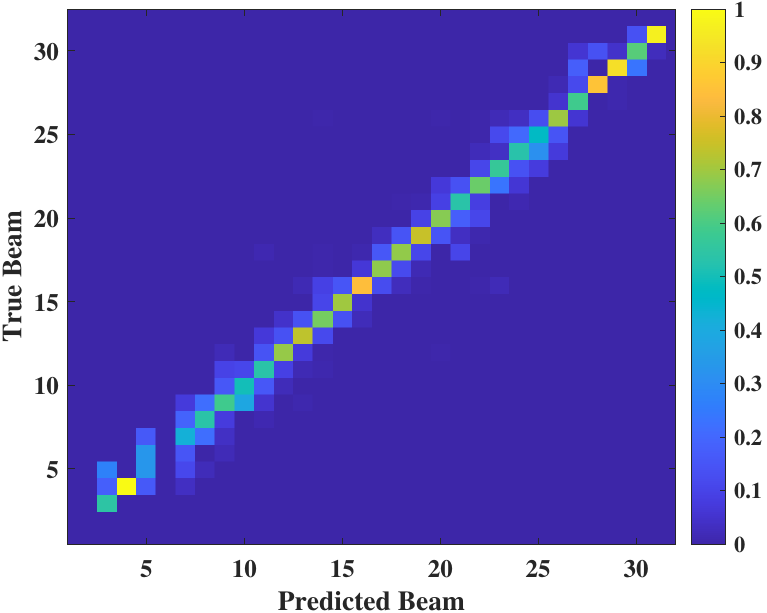}\label{fig:pos_height_dist_pred_cf}}		
		
		\caption{This figure plots the confusion matrices for the three proposed solutions: (a) vision-aided, (b) position-only, and (iii) position, height and distance combined. It is observed that by incorporating the additional height and distance information, the performance of the ML model improves significantly.    }
		\label{fig:beam_pred_cf}
	\end{figure}

	In the previous two questions, we try to answer several questions pertaining to the proposed solutions from the machine learning perspective. Next, we focus on understanding the implications of the performance of the proposed solutions from a wireless perspective. The very first thing that we explore here is the implications of the missed predictions. In Fig.~\ref{fig:beam_pred_acc}, we observe that the top-1 prediction accuracy ranges between $58 - 86\%$. This means that the percentage of missed predictions is between $14 - 42\%$, which may seem very concerning at first. Nevertheless, a closer look at the top-3 and top-5 accuracies shows that the proposed solutions can achieve close to the upper bound accuracy of $100\%$. Even the position-aided solution with the lowest prediction accuracy has a top-5 accuracy of $\approx 98\%$. The top-3 and top-5 accuracies, as shown in Fig.~\ref{fig:beam_pred_acc} highlight an important application of the sensing-aided beam prediction approach. \textbf{Instead of completely eliminating the classical beam training, the system can use the top-3 or top-5 beams to perform selective beam training, thereby reducing the overhead significantly.} This proposed combined solution can, therefore, utilize the best of both worlds to achieve the optimum beam while minimizing the training overhead. In Fig.~\ref{fig:beam_pred_cf}, we plot the confusion matrix for the three proposed solutions, i.e., vision-aided, position-alone, and position, height, and distance combined. The confusion matrix plots the true beams versus the top-1 predicted beams and shows how far the predicted values are from the ground-truth values. It highlights one important observation: Even when the ML-based solution makes a mistake, the predicted beam is nearly always within the vicinity of the ground-truth beams. This particular observation further strengthens our belief that a combined ML and wireless solution can be the most feasible and realistic alternative to traditional beam training methods.

	\subsubsection{Can the achievable receive power approach the optimal power?}
\begin{figure}[t]
		\centering
		\subfigure[Power Distribution Plot]{\centering \includegraphics[width=0.3\linewidth]{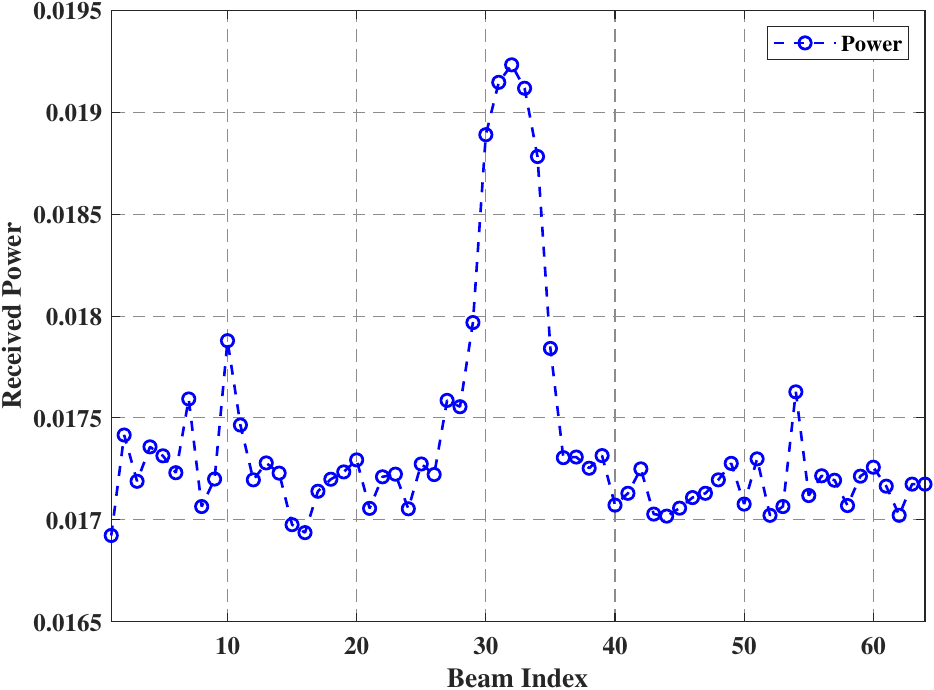}\label{fig:power_vs_beam}}	
		\subfigure[Position Beam Prediction]{\centering \includegraphics[width=0.29\linewidth]{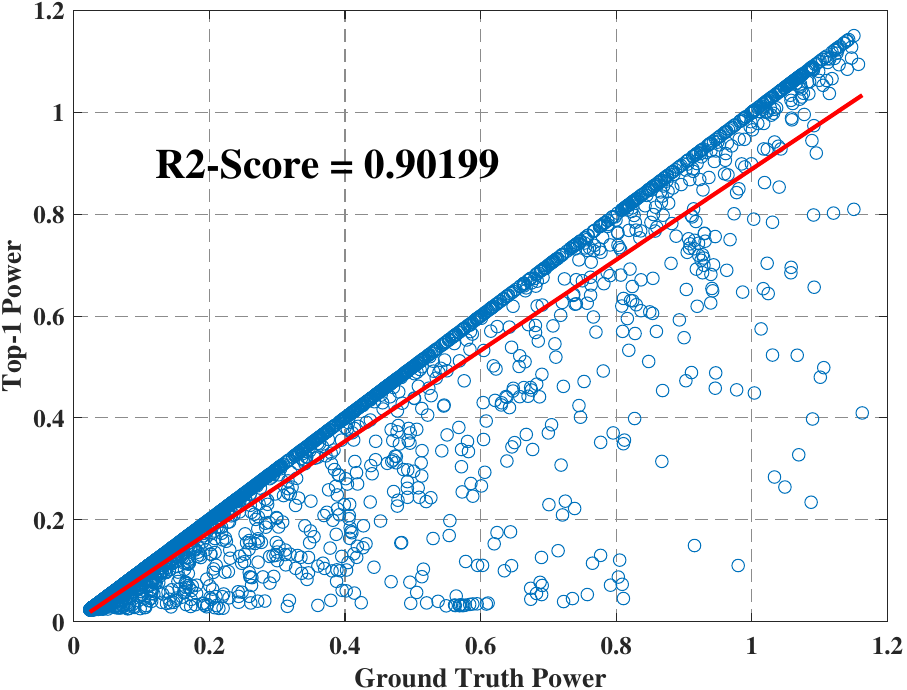}\label{fig:pos_pwr_beam_pred}}
		\subfigure[Vision Beam Prediction]{\centering \includegraphics[width=0.29\linewidth]{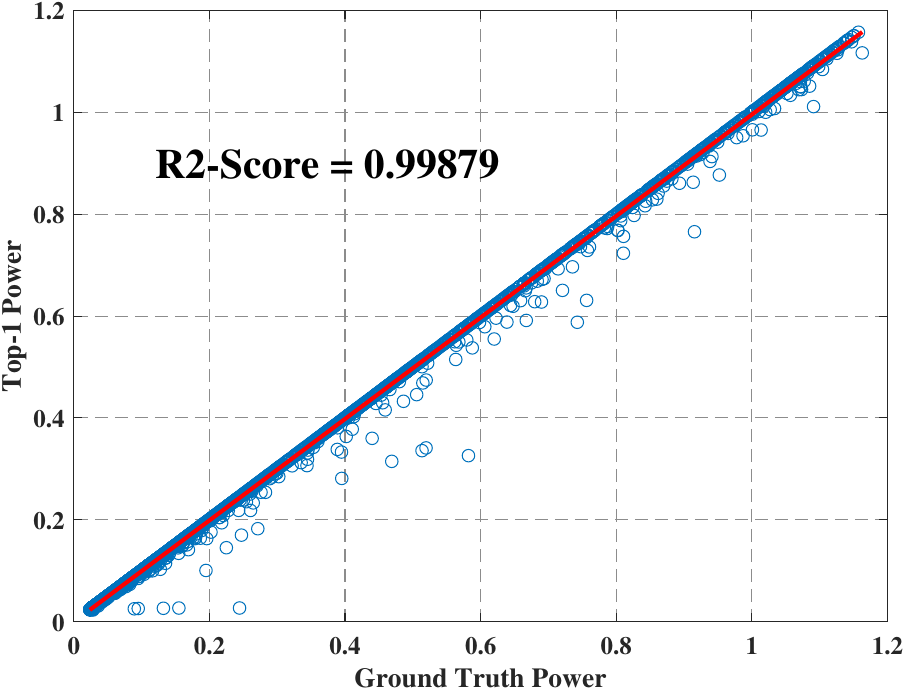}\label{fig:img_pwr_beam_pred}}

		\caption{(a) This figure plots the ground-truth receive power versus the beam indices for an example sample. In figures (b) and (c), we present the scatter plot of the top-1 receive power versus the ground-truth power and the fitted regression line (red color)) for the position-aided and vision-aided beam prediction solutions, respectively.    }
		\label{fig:power_regression_beam_pred}
	\end{figure}

	A natural extension of the previous question is studying the effect of mispredictions on the wireless system, i.e., the impact of mispredictions on the received power. Generally, only $58\%$ top-1 accuracy, as in position-aided beam prediction, will imply low received power. In Fig.~\ref{fig:pos_pwr_beam_pred} and Fig.~\ref{fig:img_pwr_beam_pred}, we present the scatter plot of top-1 received power versus the ground-truth received power and the fitted regression line (red color)) for the position-aided and vision-aided beam prediction solutions, respectively. In order to further quantify the goodness of fit, we utilize the \textit{R-squared (R2)} score. R2 score is the ratio of the variance explained by the model and the total variance and evaluates the scatter of data points across the regression line. A higher R2 score represents a better fit, i.e., a smaller difference between the observed and fitted values. For the vision-aided beam prediction, as shown in Fig.~\ref{fig:img_pwr_beam_pred}, the high beam prediction accuracy of $\approx 86\%$ is reflected in the R2-score of $0.9987$. An interesting observation here is that even with $58\%$ top-1 beam prediction accuracy, the R2-score for the position-aided beam prediction solution is $0.90$ as shown in Fig.~\ref{fig:pos_pwr_beam_pred}. \textbf{The R2-score highlights that even with the missed predictions, the received power approaches the optimal power.} The reasoning behind the high SNR, even for the missed prediction, can be found in Fig.~\ref{fig:power_vs_beam}. It presents a plot of the received power versus the beam index for one of the data samples. The figure shows that beams in the neighborhood of the optimal beams achieve similar received power to that of the ground-truth beam. We observed that the missed predictions are nearly always close to the optimal beam in the previous question itself. Therefore, it highlights that even though the top-1 prediction accuracy is low, the predicted beams are expected to achieve reasonable received power with high SNR.

	\begin{figure}[t]
		\centering
		\subfigure[Impact of Height]{\centering \includegraphics[width=0.75\linewidth]{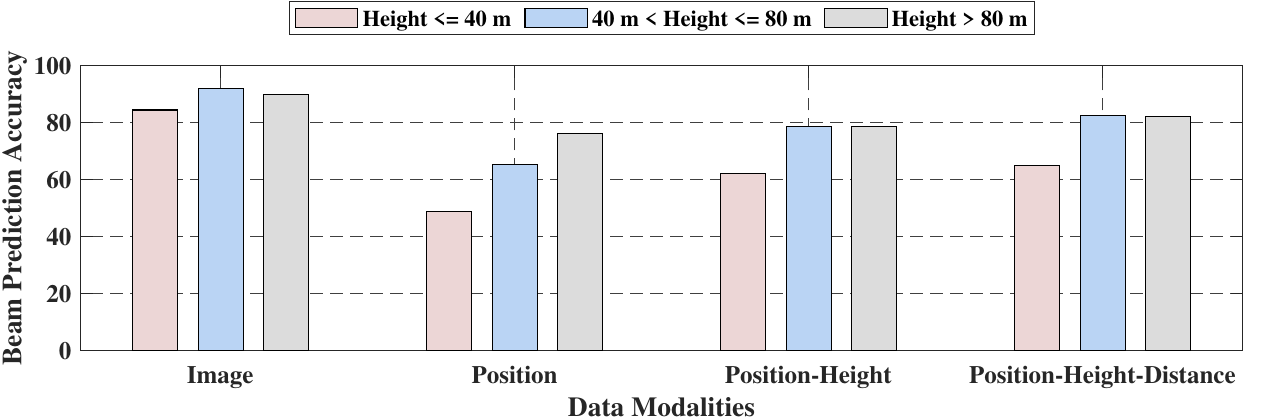}\label{fig:height_vs_acc_comp}}
		\subfigure[Impact of Speed]{\centering \includegraphics[width=0.75\linewidth]{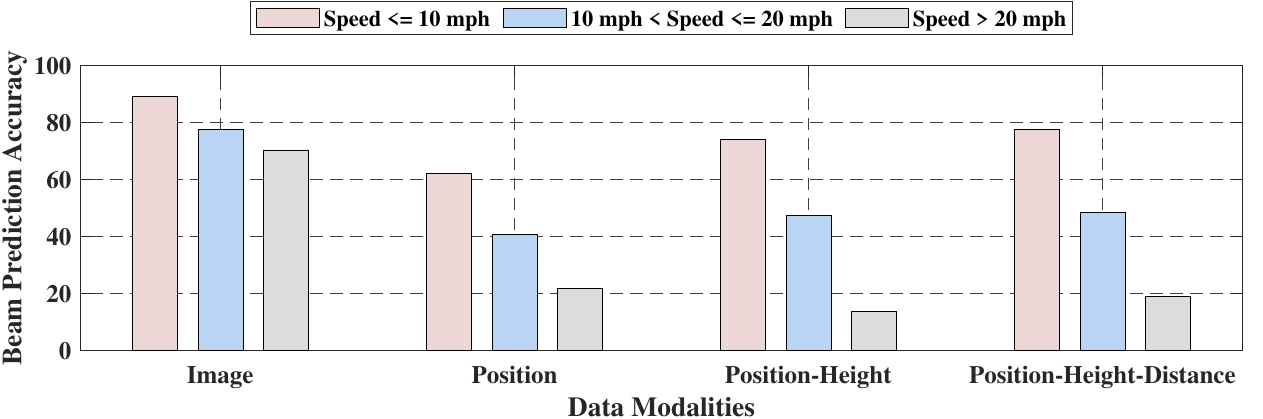}\label{fig:speed_vs_acc_com}}
		
		\caption{This figure studies the impact of (a) height and (b) speed on the beam prediction performance of the proposed solutions. It is observed that the performance is generally better at higher heights and lower speeds for all the modalities. }
		\label{fig:metric_comp}
	\end{figure}

	\subsubsection{Do the height and speed of the drone impact the beam prediction accuracy?}

	Previous sections presented a comprehensive analysis of the proposed solutions from both machine learning and wireless communication perspectives. A distinctive feature of the drone dataset is the availability of rich sensory information, including height, distance, speed, and orientation measurements. Earlier results demonstrated that incorporating additional sensing data, particularly height and distance, significantly enhanced the beam prediction accuracy compared to the position-only solution. This observation motivates a detailed investigation of how different sensing modalities impact beam prediction performance, advancing our understanding toward developing robust real-world solutions. The analysis focuses on two critical drone parameters: height and speed, examining their influence on beam prediction accuracy. The methodology involves stratifying the test dataset into three distinct sub-groups for each parameter. For height-based analysis, the data is categorized into three ranges: below 40 meters, 40-80 meters, and above 80 meters. Similarly, the speed-based analysis categorizes the data into three groups: low velocity ($\leq$ 10 mph), moderate velocity (10-20 mph), and high velocity ($>$ 20 mph).
	
	\textbf{Impact of Height on Prediction Accuracy:}
	Fig.~\ref{fig:height_vs_acc_comp} illustrates the correlation between drone height and beam prediction performance. All four approaches—image-based, position-only, position-height, and position-height-distance—exhibit reduced prediction accuracy at heights below 40 meters. This consistent observation suggests inherent challenges in beam prediction at lower altitudes, potentially attributable to increased ground reflections and multipath effects. The optimal performance range occurs between 40-80 meters across all approaches, with the image-based solution achieving approximately 90\% accuracy. The performance reduction at heights above 80 meters is less significant compared to low altitudes, though measurable. The position-based approaches demonstrate higher sensitivity to height variations compared to the image-based solution, exhibiting accuracy variations of up to 20 percentage points across different height ranges.
	
	\textbf{Impact of Speed on Prediction Accuracy:}
	The analysis in Fig.~\ref{fig:speed_vs_acc_com} reveals a significant correlation between drone velocity and prediction performance. The results indicate a consistent reduction in prediction accuracy with increasing velocity across all approaches. The image-based solution maintains higher robustness to velocity variations, sustaining accuracy above 70\% even at high velocities. In contrast, the position-based approaches exhibit substantial accuracy reduction at higher velocities, with performance decreasing to below 30\% for velocities exceeding 20 mph. This significant performance variation suggests limitations in position-based methods when handling rapid spatial changes. The position-height-distance approach, while demonstrating improved performance over the basic position-only method at low velocities (approximately 80\% versus 60\%), exhibits considerable accuracy reduction at high velocities.
	
	These quantitative findings have significant implications for practical implementation scenarios. The superior performance stability of image-based approaches under varying height and velocity conditions suggests enhanced suitability for dynamic flight operations. The identified optimal height range (40-80 meters) provides valuable parameters for flight planning and system optimization. The measured impact of high velocities on position-based methods indicates requirements for enhanced position tracking capabilities or alternative approaches to maintain accuracy during rapid drone movements.

	\section{Beam Tracking Evaluation}\label{sec:beam_track_eval}
	
	In the previous Section~\ref{sec:beam_pred_eval}, we evaluated the performance of our proposed solutions in predicting the optimal beam indices for a mmWave drone in a real-wireless environment. In this section, we shift our focus to the more challenging beam tracking task, i.e., given a sequence of sensing data, the goal is to predict up to three future beams. For this beam tracking task, we utilize the DeepSense 6G dataset (Scenario 18) as presented in Section~\ref{sec:testbed_and_dataset} and, more specifically, the sensing-aided beam tracking sequence dataset mentioned in Section~\ref{sec:task_specific_data}. In order to evaluate the performance of the proposed solution, we first present the experimental setup consisting of the neural network training hyper-parameters and the adopted evaluation metric for the beam tracking task. Next, we will discuss in detail the future beam prediction performance achieved by the three proposed approaches, i.e., beam-only, position, and vision-aided solutions. The goal of this work is to present a comprehensive evaluation of the different sensing-aided approaches, highlighting the strengths and drawbacks of the same.

	\subsection{Experimental Setup}\label{sec:beam_track_exp_setup}
	
	In this sub-section, we first present an overview of the hyper-parameters used to train the three different machine learning models as presented in Section~\ref{sec:beam_track}. Next, we discuss the metric used to evaluate the beam tracking performance of the proposed solutions. 
	
	\textbf{Network Training} The proposed solutions for the beam tracking task require a sequence of input sensing data to successfully predict the future beams. Due to the sequential nature of the input data, the preferred choice of machine learning model is the recurrent neural network. As presented in Section~\ref{sec:beam_track}, we adopt a 2-layered GRU to perform the sensing-aided beam prediction task. In order to perform a fair comparison between the three proposed solutions, we adopt the same hyper-parameters such as the number of layers, number of hidden units per layer, etc. The only difference between the model used for the beam-only approach and the other two solutions is the size of the embedding layer. Table~\ref{tab:beam_track_train_params} presents the details of the different hyper-parameters used to train and fine-tune the recurrent unit.

	%######################################################################################
	%######################################################################################
	
	\begin{table*}[!t]
		\caption{Beam Tracking: Design and Training Hyper-parameters}
		\centering
		\setlength{\tabcolsep}{5pt}
		\renewcommand{\arraystretch}{1.2}
		\begin{tabular}{@{}c|ccc@{}}
			\toprule
			\toprule
			\textbf{GRU Hyper-parameters}                & \textbf{Beam-only} & \textbf{Position} & \textbf{Image} \\ \midrule \midrule
			\textbf{Input Sequence Length}     & 8                  & 8                 & 8              \\
			\textbf{Embedding/Input Dimension} & 20                 & 2                 & 2              \\
			\textbf{Hidden State Dimension}    & 128                & 128               & 128            \\
			\textbf{Number of Output Classes}  & 32                 & 32                & 32             \\
			\textbf{Dropout Percentage}        & 0.5                & 0.5               & 0.5            \\
			\textbf{Learning Rate}             & $1 \times 10^{-3}$ & 0.01              & 0.01           \\
			\textbf{Learning Rate Decay}            & epochs 40 and 120       & epochs 40 and 120    & epochs 40 and 120   \\
			\textbf{Learning Rate Reduction Factor} & 0.1                 & 0.1                 & 0.1                \\
			\textbf{Batch Size}                & 512                & 512               & 512            \\
			\textbf{Number of Training Epochs} & 200                & 200               & 200 \\ \bottomrule \bottomrule  
		\end{tabular}
		\label{tab:beam_track_train_params}
	\end{table*}

	%######################################################################################
	%######################################################################################
	
	\textbf{Evaluation Criteria}
	Similar to the evaluation criteria used for the beam prediction task, we adopt the top-k accuracy as the metric for our beam tracking task. Since the beam tracking task involves predicting three future beams, we calculate the top-k accuracy for each of the three predicted future beams. Top-k accuracy is defined as the percentage of test samples whose ground-truth beam is within the top-k predicted beams. We evaluate the proposed solution for the beam tracking task by observing the top-1, top-2, and top-3 accuracies. It is important to highlight that the model prediction is deemed correct only if all the predicted beams match the ground-truth beams for predicting two and three future beams. For example, while evaluating the accuracy of predicting two future beams, the prediction is correct only when both the predicted values are the same as that of the two ground-truth future beams.

	\subsection{Experimental Results}
	With the experimental setup as described in Section~\ref{sec:beam_track_exp_setup}, here, we perform a comparative evaluation of the three different sensing modalities studied for the beam tracking task. The focus of this subsection is to answer some of the important questions pertaining to the feasibility of sensing-aided future beam prediction, the advantages and the limitations of different sensing modalities for the said task, etc. Similar to the beam prediction evaluation presented in Section~\ref{sec:beam_pred_results}, we evaluate the proposed solutions from both machine learning and wireless perspective. In this work, the beam tracking task entails observing a sequence of $8$ sensing data and predicting the three future beams. In order to perform a comprehensive evaluation of the proposed solution, we further extend this study to a longer sequence of 50 samples and evaluate the performance of the proposed solution on these long sequences. A detailed analysis of the proposed beam-tracking solutions is presented below.

	\subsubsection{How accurately can we predict the future beams using the proposed beam tracking solutions?}
	
\begin{figure}[t]
		\centering
		\subfigure[Future-1 Prediction Accuracy]{\centering \includegraphics[width=0.49\linewidth]{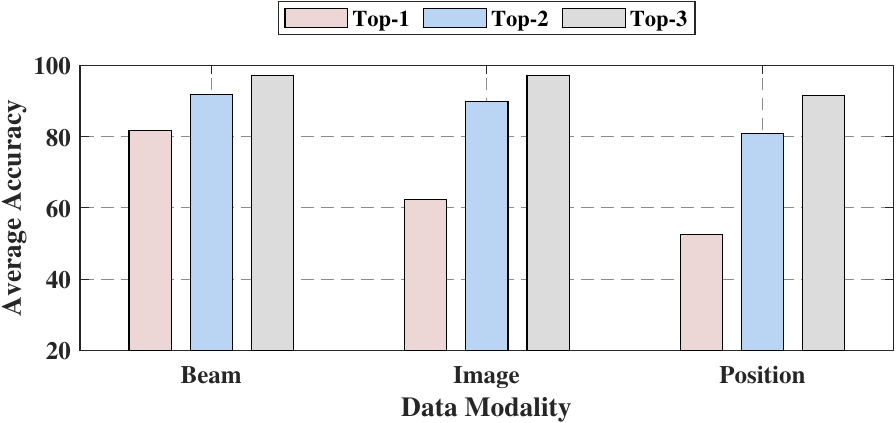}\label{fig:future1_acc}}
		\subfigure[Future-2 Prediction Accuracy]{\centering \includegraphics[width=0.49\linewidth]{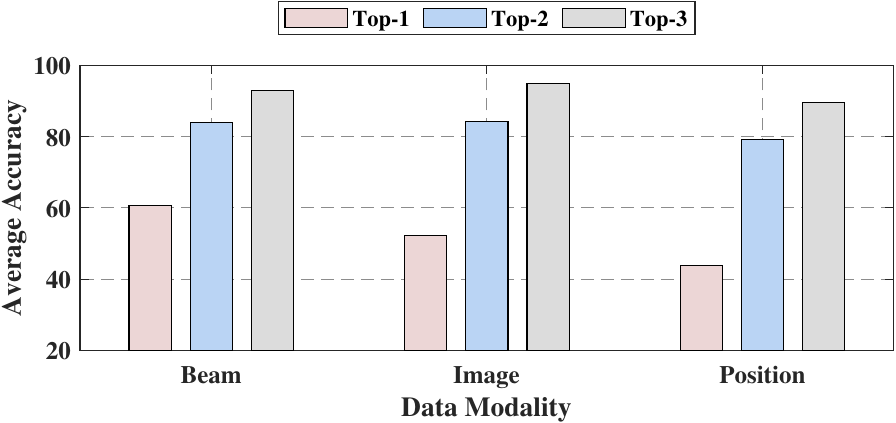}\label{fig:future2_acc}}
		\subfigure[Future-3 Prediction Accuracy]{\centering \includegraphics[width=0.49\linewidth]{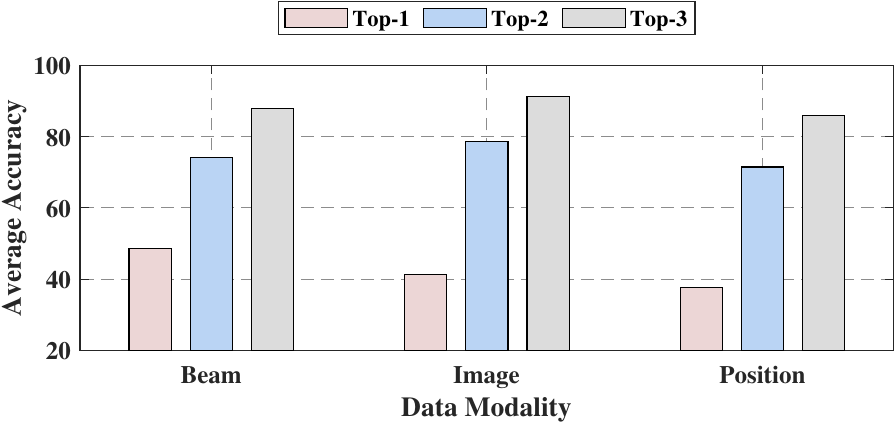}\label{fig:future3_acc}}
		
		\caption{This figure plots the top-k accuracies $(k \in (1,2,3))$ for the proposed sensing-aided beam tracking solution. Plots (a), (b), and (c) presents the accuracies achieved by the different models in predicting the first future beam, the second future beams and the third future beams, respectively. }
		\label{fig:beam_trac_acc}
	\end{figure}

	In this work, we specifically focus on three different modalities, namely, the sequence of observed beams, images, and the GPS positions of the drones, to predict the three future beams. Predicting the future beams based on the previous observations is itself a challenging task. Here, we present the future beam prediction accuracy achieved by the three proposed solutions on the test set. Fig.~\ref{fig:beam_trac_acc} shows the beam tracking accuracy achieved for all three future beams. Specifically, Fig.~\ref{fig:future1_acc}, Fig~\ref{fig:future2_acc}, and Fig.~\ref{fig:future2_acc} presents the accuracies achieved by the different models in predicting the first future beam, the second future beams and the third future beams, respectively. An interesting observation from the three figures is that the beam-based solutions achieve the highest accuracy across the three scenarios. The higher accuracy of the beam-only solution can be attributed to the following facts: (i) the beam indices are selected from a pre-defined codebook of fixed size, and (ii) by observing a sequence of previous beams, the model can easily learn to predict the future beams. The performance of the image-based solution closely follows the beam-only solution and, in fact, achieves comparable or higher top-3 accuracy for the prediction of two and three future beams. The accuracy achieved by the image-based solution highlights the potential of vision as an alternative sensing modality for this beam-tracking task. As shown in Fig.~\ref{fig:beam_trac_acc}, the position-aided solution achieves the least future beam prediction accuracy. \textbf{This further highlights the fact that position alone might not be sufficient in both beam prediction and beam tracking tasks.} The lower accuracy of the position-aided solution highlights the need for additional sensing data, such as the height and distance of the drone relative to the base station.

	The other key observation in the above figures is that the prediction accuracy decreases for all three data modalities as we predict further into the future. For example, the beam-alone solution predicts the two and three future beams with $\approx 92\%$ and $\approx 88\%$ accuracies (considering the top-3 accuracy), whereas the accuracy for predicting only one future beam is $\approx 98\%$. The decrease in accuracy calls attention to the increasing difficulty in predicting beams that are further in the future. Compared to the beam-alone approach, the decrease in accuracy from predicting one future beam to three future beams is relatively small for the vision-aided and position-aided solution. The decrease in accuracy for the beam-alone approach is $10.20\%$, whereas, for the vision-aided and position-aided solutions, the decrease in accuracy is $6.03\%$ and $6.15\%$, respectively. Although the beam-alone approach achieves the best beam tracking accuracy, the decrease in accuracy as we try to predict further in the future highlights the limitations of the said approach and the need for exploring other sensing modalities, such as, in this case, visual and positional data.

	\subsubsection{What are the implications of missed prediction?}
\begin{figure}[t]
		\centering
		\subfigure[Beam Only]{\centering \includegraphics[width=0.32\columnwidth]{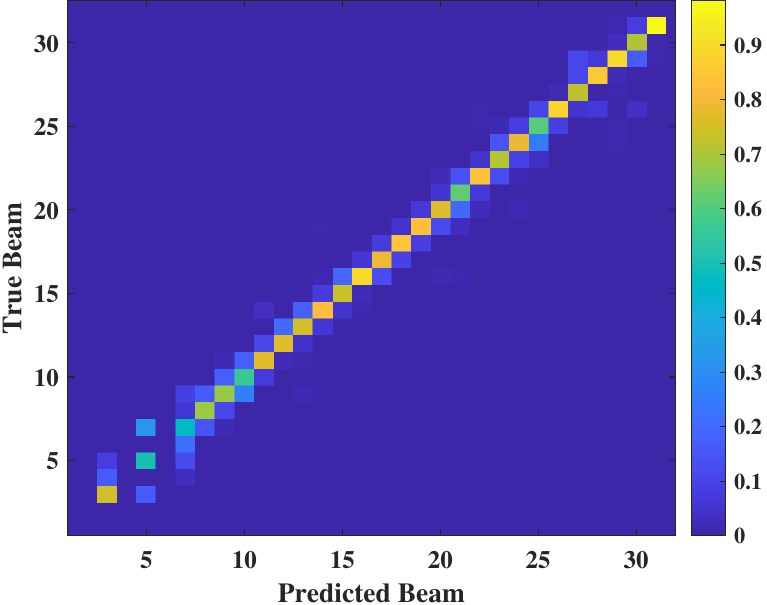}\label{fig:beam_only_track_cf}}
		\subfigure[Vision]{\centering \includegraphics[width=0.32\columnwidth]{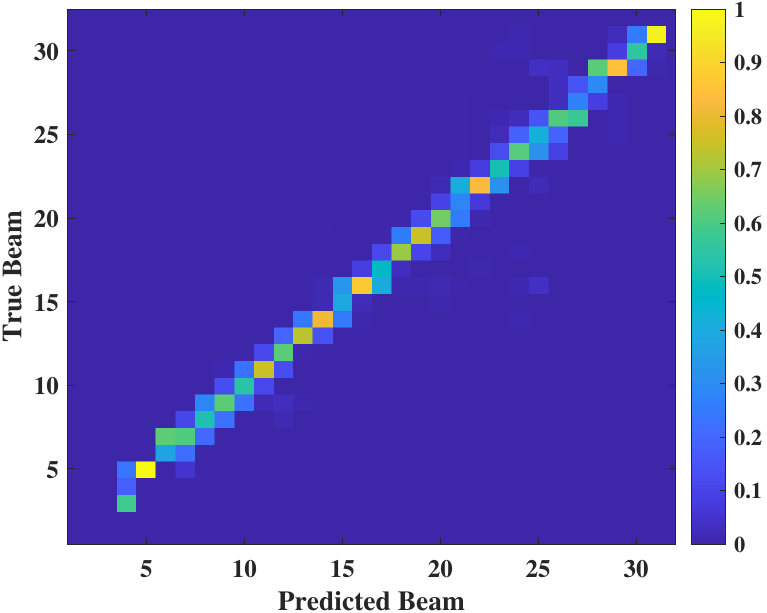}\label{fig:img_only_track_cf}}
		\subfigure[Position]{\centering \includegraphics[width=0.32\columnwidth]{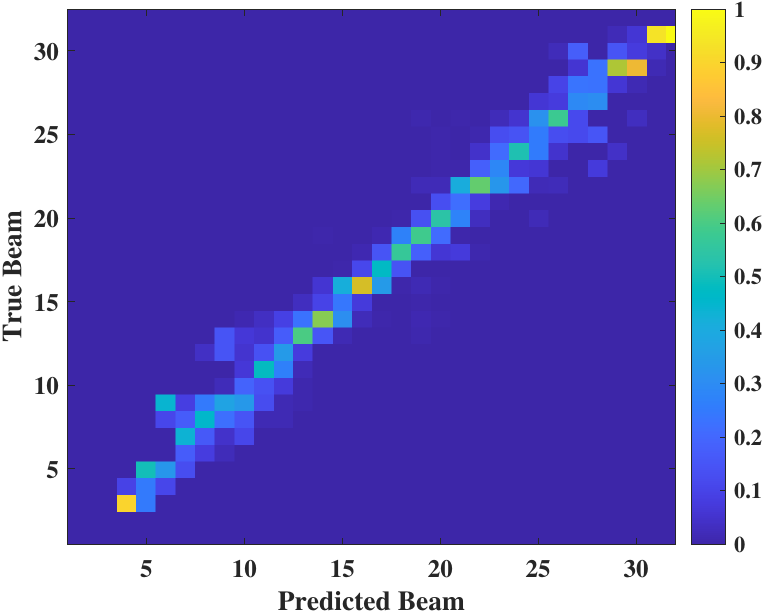}\label{fig:pos_only_track_cf}}	
		\subfigure[Training Dataset Distribution]{\centering \includegraphics[width=0.45\columnwidth]{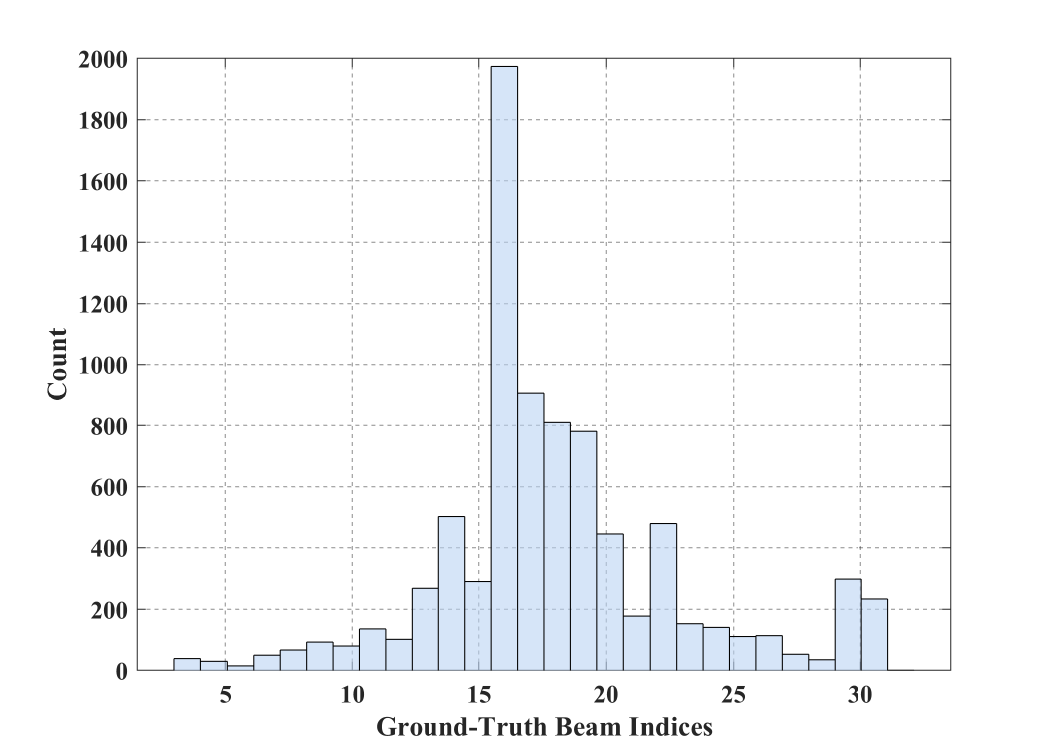}\label{fig:seq_train_data_hist}}
		\subfigure[Test Dataset Distribution]{\centering \includegraphics[width=0.45\columnwidth]{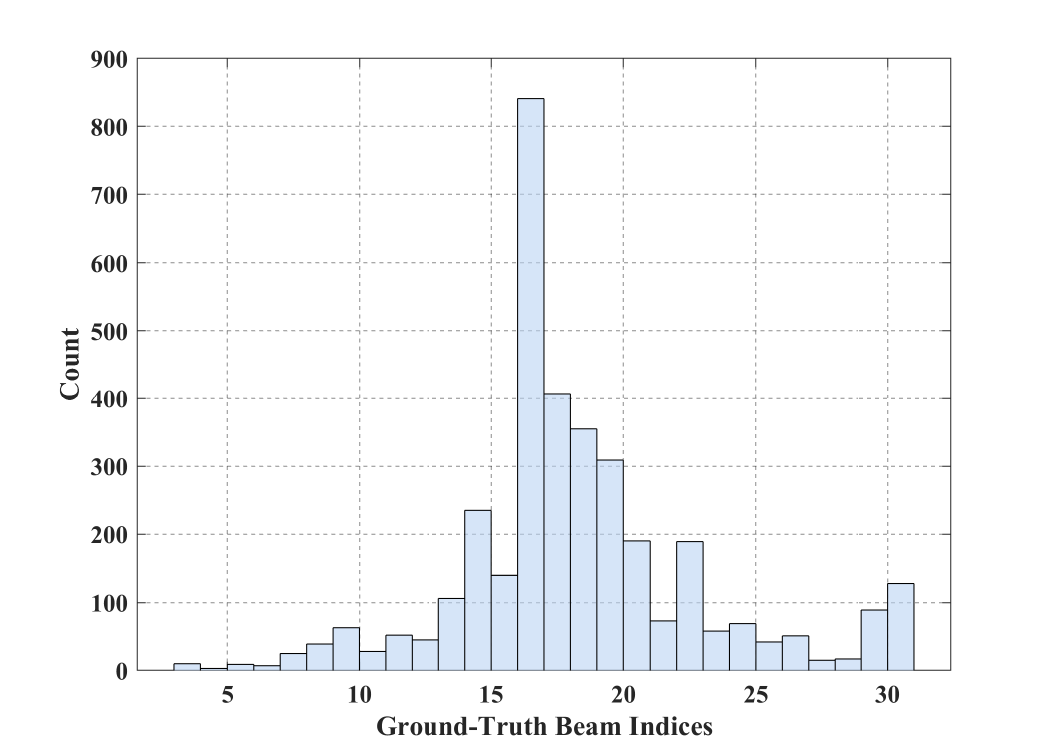}\label{fig:seq_test_data_hist}}	
		\caption{Figures (a), (b) and (c) plots the confusion matrices of the top-1 predicted beams for all beam-only, vision-based and position-based approaches. In (d) and (e), we preset the number of samples per beam index in the training and test dataset.   }
		\label{fig:cf_beam_tracking}
	\end{figure}

	In Fig.~\ref{fig:beam_trac_acc}, we present the top-1, top-2, and top-3 accuracies for predicting up to three future beams. The top-1 accuracies achieved by the three proposed solutions for predicting just one future beam are $81.62\%$, $62.28\%$, and $52.54\%$. The low accuracy numbers would point towards a significantly large number of mistakes in predicting the future beam. The concern here is more from the wireless perspective since wrong predictions can result in low SNR and defeat the entire purpose of this beam-tracking task. In order to verify the impact of the mispredictions, we plot the confusion matrix that summarizes the prediction results by comparing the predicted beams with the ground-truth beams. In Fig.~\ref{fig:beam_only_track_cf}, Fig.~\ref{fig:img_only_track_cf}, and Fig.~\ref{fig:pos_only_track_cf}, we plot the confusion matrices of the top-1 predicted beams for all beam-only, vision-based and position-based approaches. The important observation from all three confusion matrices is that even for the mispredictions, the predicted beam is always in the neighborhood of the optimal beam indices. Therefore, the confusion matrix indicates that the predicted future beams will lead to high SNR even when the prediction is off by a small margin. Another interesting observation here is that for both beam-only and image-based approaches, the future beam prediction accuracy is considerably lower for beam indices between $1 - 10$. The lower accuracy for beam indices between $1 - 10$ can primarily be attributed to the distribution of samples per beam across the dataset. In Fig.~\ref{fig:seq_train_data_hist} and Fig.~\ref{fig:seq_test_data_hist}, we preset the number of samples per beam index in the training and test dataset. As observed in the two figures, the number of samples per beam index is not uniform for both training and test datasets. In fact, for the beam indices between $1 - 10$, the numbers are specifically low, with less than ten samples for some beam indices in the training data. The data imbalance further highlights that additional data can help in improving the performance even further.

	\subsubsection{Are the sequences of previously observed beam indices enough to predict future beams perpetually?}
	
	In this work, the beam tracking task entails predicting up to three future beams based on a sequence of eight previously observed sensing data. In the previous two discussions, we observed that the beam-alone approach obtains the best future beam prediction accuracy for all three scenarios predicting one, two, and three future beams. Although the performance image-based solutions follow closely that of the beam-alone approach, there is a significant difference in the prediction accuracy. For instance, the top-1 accuracy of predicting one future beam for the beam-alone approach is $\approx 82\%$, whereas the image-based solution only achieves $\approx 62\%$ beam tracking accuracy. However, an important observation here is that the higher prediction accuracy of the beam-only approach comes at a significant cost for beam training overhead. As mentioned earlier in Section~\ref{sec:baseline_beam_track}, in the beam-alone approach, the proposed ML model is trained on a sequence of $8$ input previously observed ground-truth beam values to predict up to three future beams. The only process of obtaining the ground-truth beam indices is by performing beam training, which is typically associated with a large training overhead. One of the primary motivations of this work is to propose an alternative solution to reduce and minimize this beam training overhead. In order to reduce the beam training overhead for the beam-alone approach, an alternative solution is to utilize the latest predicted beam index as an input to predict the next future beam. In essence, this would indicate that we only need to perform the beam training to obtain the initial $8$ ground-truth beam indices, and the ML model will be able to predict the future beam indices. Therefore, the critical question here is that \textbf{ can the beam-alone approach achieve similar future beam prediction accuracy without the need for repetitive beam training? } 
	
\begin{figure}[t]
		\centering
		\subfigure[]{\centering \includegraphics[width=.48\columnwidth]{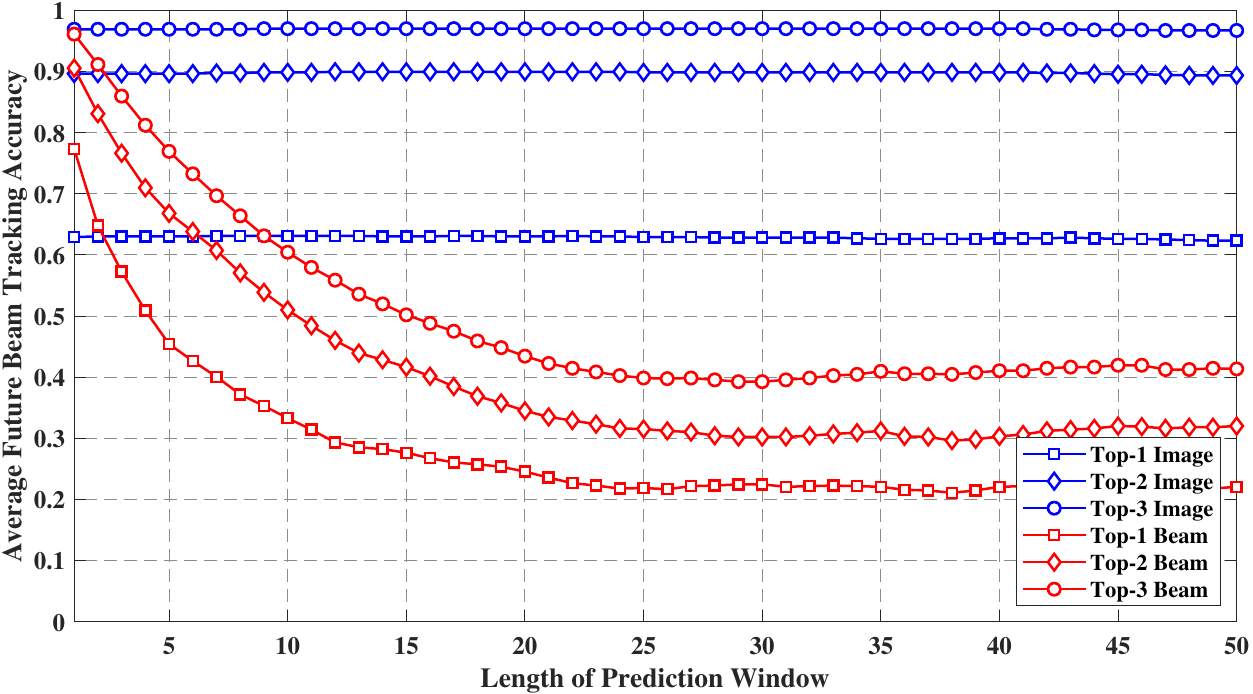}\label{fig:imv_bs_beam_seq_50}}
		\subfigure[]{\centering \includegraphics[width=.48\columnwidth]{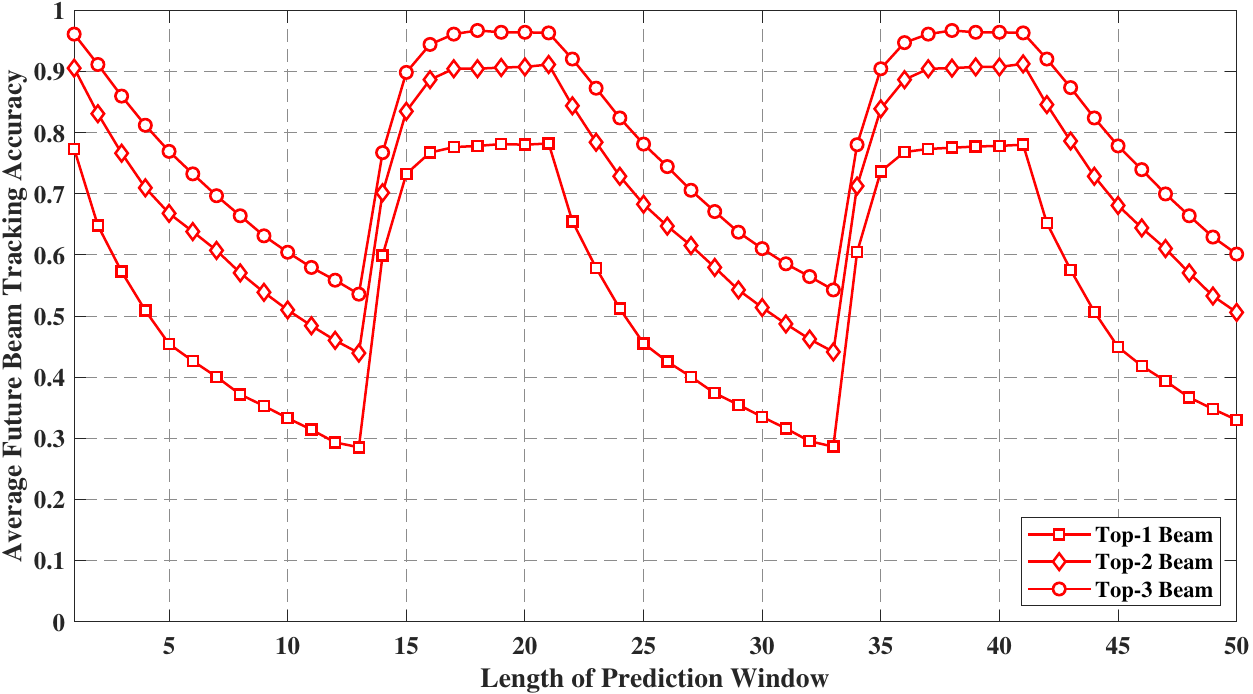}\label{fig:imv_bs_beam_seq_50_interval}}		
		%\subfigure[]{\centering \includegraphics[width=0.49\columnwidth]{figures/img_vs_beam_resource_utilization.pdf}\label{fig:imv_bs_beam_resource_util}}		
		\caption{In (a), we present the top-1, 2 and 3 beam prediction accuracies of the beam-alone and vision-aided approach on long-sequences of 50 samples. The performance of the beam-only approach reduces on average by $\approx 75\%$, highlighting the challenges of relying only on past beams to predict future beam for a longer period. Figure (b), presents the future beam prediction accuracy with intermittent beam training for the beam-only approach. }
		\label{fig:lon_seq}
	\end{figure}

	In order to answer this question, we experiment with a small dataset ($1005$ data samples) consisting of long sequences of consecutive $50$ samples where the beam training is done for only the first $8$ samples. In the first time step, the $8$ ground-truth beams are utilized as an input to predict one future beam. At the next step, the previously predicted beam forms the $8$th sample in the input sequence and is provided to the ML model. Therefore, after eight steps, the input sequence consists of only the predicted beams, and the entire process is repeated 50 times in this experiment. We compare the performance of the beam-alone solution in this experiment with the vision-aided proposed approach. The input to the vision-aided ML models is the sequence of $8$ previously captured RGB images for all the $50$-time steps. In Fig.~\ref{fig:imv_bs_beam_seq_50}, we present the performance achieved by both the beam-alone and vision-aided approach on these long sequences. As observed, the beam-alone approach achieves better accuracy for the initial few samples. Nevertheless, as we try to predict further into the future, the performance of the beam-alone severely degrades, with the top-3 accuracy dropping from $\approx 97\%$ to $\approx 40\%$. For the beam-alone scenario, as the error in the predicted value accumulates, the predicted beam value diverges from the ground-truth beam indices, resulting in this considerable drop in beam tracking accuracy. On the other hand, the vision-aided solution achieves a consistent performance in predicting the future beam with top-1 and top-3 accuracies of $\approx 62\%$ and $\approx 97\%$, respectively. The vision-aided solution consistently outperforms the beam-alone solution when the prediction window is longer than $3$ or $4$ time steps. It is important to highlight here that the vision-aided solution does not require any beam training, and therefore, there is no overhead to the wireless system associated with this approach. 
	
		\begin{figure}[!t]
		\centering
		\includegraphics[width=0.75\linewidth]{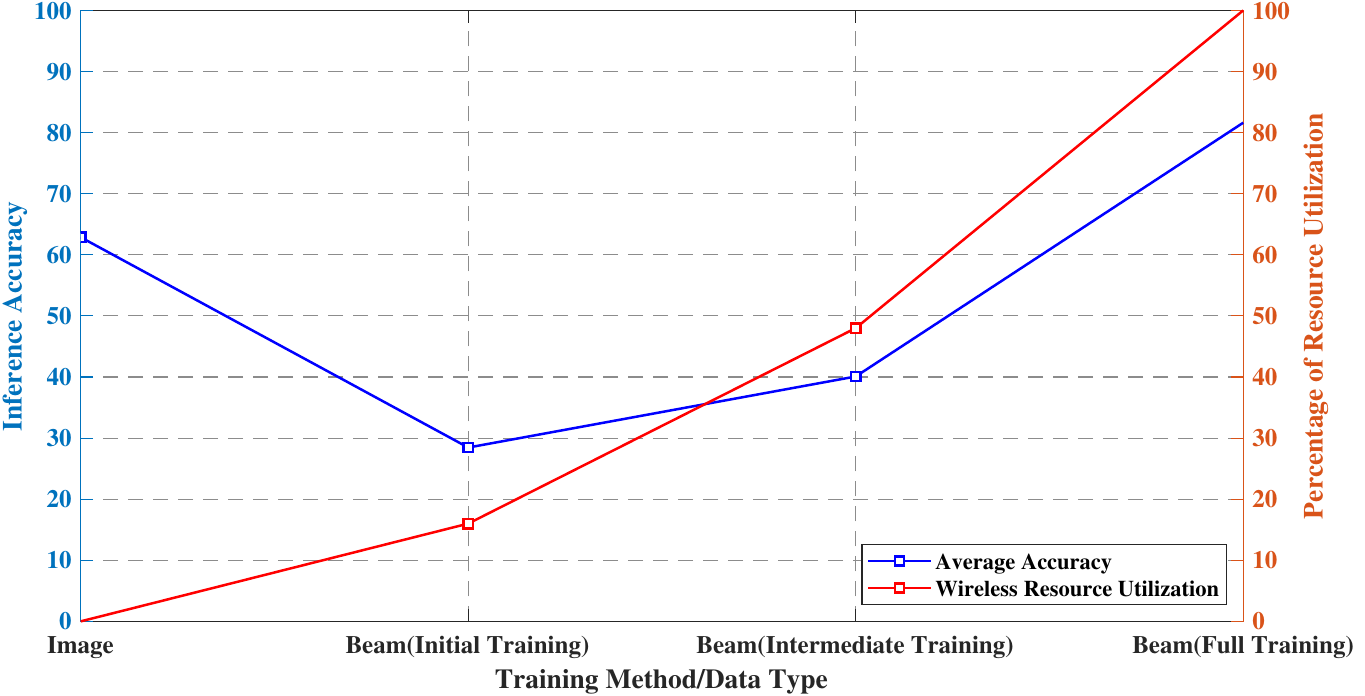}
		\caption{This figure plots the future beam tracking accuracy and the percentage of wireless resources needed for all four approaches discussed in this work.   }
		\label{fig:imv_bs_beam_resource_util}
	\end{figure}

	In this particular experiment, we only perform beam training for the initial $8$ inputs for the beam-alone approach. As observed in Fig.~\ref{fig:imv_bs_beam_seq_50}, the beam-alone approach suffers from significant performance degradation when predicting for this long sequence of 50-time steps. In order to further evaluate the performance of the beam-alone approach, we propose a modification to the experiment. The modifications entail that instead of just performing beam training initially, we repeat the beam training intermittently. Fig.~\ref{fig:imv_bs_beam_seq_50_interval} presents the future beam prediction accuracy with intermittent beam training for the beam-only approach. For this modified experiment, we perform beam training from time-steps $13-20$ and again from $33-40$. We observe that intermittent beam training does help to boost the performance of the beam-alone approach and achieve the optimum prediction accuracy. However, the effect of the beam training lasts for a small window of future prediction, and the accuracy starts degrading again. Even with intermittent training, the low future beam prediction accuracy highlights the shortcomings and limitations of the beam-alone approach. It highlights the significant advantage of relying on other sensing modalities, such as visual data, to predict future beams.

	\subsubsection{What is the trade-off between wireless communication resources and accuracy for the three proposed approaches?}
	
	In this discussion, we propose to study the trade-off between the percentage of wireless communication resources needed and the beam tracking accuracy of the proposed solution. This work refers to the beam training overhead as the wireless communication resource needed. For this study, we consider the long sequence of $50$ samples as presented in the previous discussion and compare the following four different approaches: (i) vision-based beam tracking, (ii) beam-only approach with just initial beam training of $8$ sequences, (iii) beam-only approach with intermittent beam training, and (iv) beam-alone approach with beam training at every time step. The percentage of wireless resources needed is calculated based on the number of beam training required out of the $50$ time-steps in this particular experiment. Fig.~\ref{fig:imv_bs_beam_resource_util} plots both the future beam tracking accuracy and the percentage of wireless resources needed for all four approaches. As observed here, the vision-based solution does not require any beam training, and hence, the percentage of wireless resources needed is zero. The beam-alone approach with beam training for every time step achieves the best accuracy but utilizes $100\%$ of the resources. The other two approaches utilize lower wireless resources but at the cost of beam tracking accuracy.

	\section{Conclusion}
	In this study, we introduce a novel strategy that utilizes sensory data, including visual and positional information, to enhance beam prediction in millimeter-wave (mmWave) and terahertz (THz) drone communication systems, leveraging real-world scenarios from the DeepSense 6G dataset. Our in-depth evaluation showcases significant advantages of this sensory data integration, with the vision-aided method achieving top-1 and top-5 accuracies of 86.32\% and 99.69\%, respectively, thereby substantially reducing beam training overhead. This approach not only improves beam prediction accuracy at the current time step but also exhibits promising capabilities in predicting future beam alignments, addressing the dynamic nature of drone mobility and the associated real-time processing challenges. Through this pioneering work, we lay a foundational step towards realizing efficient, highly-mobile 6G drone communications, illustrating the critical role of advanced sensory data in enhancing future aerial networks. In future work, it is interesting to explore how other sensing modalities, such as radar, could be integrated and fused with visual and location data. Further, it is interesting to investigate how digital twins, \cite{jiang2024learnablewirelessdigitaltwins}, could be leveraged to provide further perception to drone communication networks and potentially reduce the require training datasets.

	%==========
	\balance

	%\bibliography{Ref_fina }

\end{document}